\documentclass[11pt,a4paper]{article}
\usepackage{jheppub}
\usepackage[T1]{fontenc}
\usepackage{amssymb}
\usepackage{mathtools}
\usepackage{stackengine}
\usepackage{scalerel}

\usepackage{hyperref}
\usepackage{epsfig}
\usepackage{amsmath,latexsym,amssymb}
\usepackage{graphicx}
\usepackage{braket}
\usepackage{pdfsync}

\usepackage{framed}
\usepackage{mathtools}

\usepackage{jheppub}
\usepackage[T1]{fontenc}
\usepackage{amssymb}
\usepackage{mathtools}
\usepackage{amsmath}
\usepackage{bm}
\usepackage{stackengine}

\usepackage{scalerel}
\usepackage{eufrak}
\usepackage{framed}

\usepackage{mathtools}

\usepackage{pstricks}

\usepackage{axodraw4j}
\usepackage{color}
\usepackage{pstricks}
\usepackage{amssymb}

\usepackage{amsmath}
\usepackage{bbold}
\usepackage{bm}
\usepackage{latexsym}
\usepackage{braket}
\usepackage{slashed}
\usepackage{graphicx,booktabs,multirow}
\usepackage{eufrak}
\numberwithin{equation}{section}

\usepackage{color}
\usepackage{pstricks}
\usepackage{amssymb}

\makeatletter
\newcommand{\doublewidetilde}[1]{{%
  \mathpalette\double@widetilde{#1}%
}}
\newcommand{\double@widetilde}[2]{%
  \sbox\z@{$\m@th#1\widetilde{#2}$}%
  \ht\z@=.9\ht\z@
  \widetilde{\box\z@}%
}
\makeatother

\usepackage{hyperref}
\usepackage{slashed}
\usepackage{epsfig}
\usepackage{amsmath,latexsym,amssymb}
\usepackage{graphicx}
\usepackage[latin1]{inputenc}
\usepackage{braket}
\usepackage{pdfsync}

\usepackage{framed}

\usepackage{mathtools}

\def\be{\begin{equation}}
\def\ee{\end{equation}}
\def\ba{\begin{eqnarray}}
\def\ea{\end{eqnarray}}

\newcommand{\bz}{\bar{z}}

\newcommand{\bh}{\bar{h}}

\newcommand{\bx}{\bar{x}}

\newcommand{\zbar}{\bar{z}}

\newcommand{\hy}{{}_{2}F_1}

\def\D{\Delta}

\newcommand{\comment}[1]{}

\newcommand{\eea}{\end{eqnarray}}

\author{
Wei Fan${}^{1}$, Angelos Fotopoulos${}^{2}$, Stephan Stieberger${}^{3}$,
Tomasz R.\ Taylor${}^{2}$,\, Bin Zhu${}^2$\\[0.5cm] $^1$\it
Department of Physics, School of Science, Jiangsu University of Science and Technology,
Zhenjiang, 212003, China\\[0.2cm]
 $^2${\it Department of Physics \\
  Northeastern University, Boston, MA 02115, USA}\\[0.2cm]
$^3${\it Max--Planck--Institut f\"{u}r Physik,	Werner--Heisenberg--Institut, \\80805 M\"unchen, Germany}}
\emailAdd{fanwei@just.edu.cn}
\emailAdd{a.fotopoulos@northeastern.edu}
\emailAdd{stephan.stieberger@mpp.mpg.de}
\emailAdd{taylor@neu.edu}
\emailAdd{zhu.bi@northeastern.edu}

\title{\boldmath Elements of Celestial Conformal Field Theory \unboldmath}

\abstract{In celestial holography, four-dimensional scattering amplitudes are considered as two-dimensional conformal correlators of a putative two-dimensional celestial conformal field theory (CCFT). The simplest way of converting momentum space amplitudes into CCFT correlators is by taking their Mellin transforms with respect to light-cone energies. For massless particles, like gluons, however, such a construction leads to three-point and four-point correlators that vanish everywhere except for a measure zero  hypersurface of celestial coordinates. This is due to the four-dimensional momentum conservation law  that constrains the insertion points of the operators associated with massless particles. These correlators are reminiscent of Coulomb gas correlators that,  in the absence of background charges, vanish due to charge conservation. We supply the background momentum by coupling Yang-Mills theory to a background dilaton field, with the (complex) dilaton source localized on the celestial sphere.  This picture emerges from the physical interpretation of the solutions of the system of differential equations discovered by Banerjee and Ghosh. We show that the solutions can be written as Mellin transforms of the amplitudes evaluated in such a dilaton background.  The resultant three-gluon and four-gluon amplitudes are single-valued functions of celestial coordinates enjoying crossing symmetry and all other properties expected from standard CFT correlators. We use them to extract OPEs and compare them with
the OPEs extracted from multi-gluon celestial amplitudes without a dilaton background.
The leading OPE terms with single holomorphic poles are identical, but the terms with antiholomorphic poles are replaced by the terms involving ``quasishadow'' fields.
We perform the conformal block decomposition of the four-gluon single-valued correlator and determine the dimensions, spin and group representations of the entire primary field spectrum of the Yang-Mills sector of CCFT.}


\begin{document}
\maketitle
\section{Introduction}
The geometry of the four-dimensional asymptotically flat world is described by Lorentzian metrics. The universal cover of the Lorentz symmetry group is isomorphic to the $SL(2,C)$ group of conformal transformations in two dimensions. In celestial holography, $SL(2,C)$ acts on the celestial sphere and is extended to the infinite-dimensional group of BMS superrotations \cite{Strominger:2017zoo,Raclariu:2021zjz,Pasterski:2021rjz}. Four-dimensional physics, in particular the scattering processes, are described in the framework of a putative, two-dimensional celestial conformal field theory (CCFT) with the Virasoro operators identified as the generators of superrotations.

Pseudo-Riemannian geometry with a Lorentzian signature endows  the celestial sphere with the complex coordinate $z$ and its complex conjugate $\bz=z^*$. The tools of complex analysis, in particular Cauchy theorem and analytic continuation, are indispensable to two-dimensional conformal field theories
\cite{DiF}, thus it is expected that they will be equally useful in CCFT. In CFT, the form of two- and three-point correlation functions of primary fields are determined by conformal invariance. In many theories, including minimal models and in the WZW model,  the four-point functions can be constructed from the solutions of partial differential equations by imposing physically sensible monodromy properties \cite{DiF,dots}.

{}For the scattering amplitudes, the currently adopted holographic dictionary is very simple. With each incoming or outgoing particle, there is an associated primary field of dimension $\Delta$ and two-dimensional spin $J$ equal to its helicity. The correlators of such primary fields, also known as  celestial amplitudes, are identified with the Mellin transforms of the momentum space amplitudes with respect to the energies of external particles
\cite{Strominger1706,Pasterski1705}. The dimensions $\Delta$ are dual to energies in the Mellin sense. For massless particles propagating at the speed of light, the operator insertion points $z\in C$ determine their asymptotic momentum directions. But this cannot be the end of the story. The amplitudes are evaluated on-shell, with the momenta subject to the conservation law, therefore the operator insertion points are constrained. As a result, for a generic configuration of the operator insertion points, all three-point and four-point correlators of massless particles are zero due to such kinematic constraints. In the case of four particles though, the correlators have non-vanishing support on a measure zero hypersurface of coordinates with real cross ratios characterizing planar scattering events. They can only be written as distribution-valued functions of complex variables \cite{Strominger1706}. The vanishing of such correlation functions is a major stumbling block that prevents us from formulating CCFT along the lines of standard textbooks.

A similar problem appears in the Coulomb gas formulation of minimal models: the correlators vanish in the absence of background charges \cite{DiF,dots}. {}From the above discussion, it is clear that celestial amplitudes are missing some background momentum ``charges.'' In this work, we consider pure gluon amplitudes and supply external momentum by coupling Yang-Mills theory to the background field of a complex dilaton, in a ``minimal'' way, akin to the universal dilaton/axion-gauge boson couplings in heterotic superstring theory. The source of dilaton is localized on celestial sphere, therefore the dilaton ``charge'' plays the same role in CCFT as the $U(1)$ background charge in Coulomb gas models.
To reach this conclusion, we only need to assume
that four-point celestial amplitudes satisfy the system of partial differential equations recently discovered by
Banerjee and Ghosh \cite{Banerjee:2020vnt}.

This paper is organized as follows. In Section 2, we reorganize and solve the Banerjee-Ghosh (BG) equations for four-point celestial amplitudes.
The solutions, hereafter called single-valued celestial amplitudes, are well defined on the entire complex plane of the cross ratio. Their conformally soft limit is related to the correlator constructed (at the end of a very indirect route) in Ref.\cite{II}. In Section 3, we introduce the dilaton background on the celestial sphere and couple it to Yang-Mills theory. We compute three-gluon and four-gluon correlators in the presence of such a background and show that they are identical to the solutions of BG equations. In Section 4, we extract the leading order OPEs from single-valued celestial amplitudes and compare them with the OPEs of Refs.\cite{Fan1903,Strominger1910}, extracted from  Mellin-transformed multi-gluon amplitudes evaluated without a dilaton  background. The terms with single holomorphic poles are identical, but the terms with antiholomorphic poles are replaced by the terms involving ``quasishadow'' fields. We attribute this deformation to certain ``MHV projection.'' In Section 5,
we perform the conformal block decomposition of the four-gluon correlators along the lines of Refs.\cite{I,II}. We determine the dimensions, spin and group representations of the entire primary field spectrum of the Yang-Mills sector of CCFT.

\section{Solutions of Banerjee-Ghosh equations}
The physical content of BG equations for Mellin-transformed celestial amplitudes \cite{Banerjee:2020vnt} is the same as of the equations satisfied by singular vectors in minimal models and of Knizhnik-Zamolodchikov equations in WZW models. They all describe decoupling of certain descendants of primary fields. BG equations follow from the consistency requirements of the subleading soft theorem with the OPEs of primary fields associated with gluons. They rely on the subleading soft behavior of MHV amplitudes. This is completely sufficient for four-gluon amplitudes which are non-vanishing only for the MHV configurations of external gluons.

The four-gluon celestial MHV amplitude can be expressed in terms of two partial amplitudes \cite{Taylor:2017sph}, $M(z_i,\bz_i)$ and $\widetilde M(z_i,\bz_i)$:
\begin{align}
\Big\langle\phi_{\D_1,-}^{a_1,-\epsilon}(z_{1},\zbar_{1}
)\,
\phi_{\D_2,-}^{a_2,-\epsilon}&(z_2,\bz_2)
\,\phi_{\D_3,+}^{a_3,+\epsilon}
(z_3,\bar z_3)
\,\phi_{\D_4,+}^{a_4,+\epsilon}(z_4,\bz_4)\Big\rangle= \nonumber\\[1mm] &~~=f^{a_1a_2b}f^{a_3a_4b}M(z_i,\bz_i)
+f^{a_1a_3b}f^{a_2a_4b} \widetilde M(z_i,\bz_i) .\label{partials}
\end{align}
Here, the subscripts $\Delta_i,J_i=\pm 1$ refer to the dimensions and spins (gluon helicities) of the primary fields, respectively, group indices are labeled by $a_i$, while the superscripts $\pm\epsilon$ indicate incoming
$(-\epsilon)$ and outgoing $(+\epsilon)$ particles. BG equations  \cite{Banerjee:2020vnt} are equivalent to two sets of four partial differential equations for each partial amplitude \cite{Hu:2021lrx}:
\begin{align}
&\Big( \partial_4-\frac{\Delta_4}{z_{34}}-\frac{1}{z_{14}}+ \frac{\Delta_{3}-J_{3}-1+\bar{z}_{34}\bar{\partial}_{3}}{z_{34}} e^{\frac{\partial}{\partial{\Delta_4}}-\frac{\partial}{\partial{\Delta_{3}}}}\Big)M(z_i,\bz_i) = 0\ , \label{eq:A1_1} \\
&\Big( \partial_3-\frac{\Delta_3}{z_{43}}-\frac{1}{z_{23}} + \frac{\Delta_{4}-J_{4}-1+\bar{z}_{43}\bar{\partial}_{4}}{z_{43}} e^{\frac{\partial}{\partial{\Delta_3}}-\frac{\partial}{\partial{\Delta_{4}}}}\Big) M(z_i,\bz_i) = 0\ , \label{eq:A1_2} \\
&\Big( \partial_3-\frac{\Delta_3}{z_{23}}-\frac{1}{z_{43}}- \frac{\Delta_{2}-J_{2}-1+\bar{z}_{23}\bar{\partial}_{2}}{z_{23}} e^{\frac{\partial}{\partial{\Delta_3}}-\frac{\partial}{\partial{\Delta_{2}}}}\Big) M(z_i,\bz_i) = 0\ , \label{eq:A1_3} \\
&\Big( \partial_4-\frac{\Delta_4}{z_{14}}-\frac{1}{z_{34}} - \frac{\Delta_{1}-J_{1}-1+\bar{z}_{14}\bar{\partial}_{1}}{z_{14}} e^{\frac{\partial}{\partial{\Delta_4}}-\frac{\partial}{\partial{\Delta_{1}}}}\Big)M(z_i,\bz_i) = 0\ , \label{eq:A1_4}
\end{align} and
\begin{align}
&\Big( \partial_4-\frac{\Delta_4}{z_{24}}-\frac{1}{z_{14}} - \frac{\Delta_{2}-J_{2}-1+\bar{z}_{24}\bar{\partial}_{2}}{z_{24}} e^{\frac{\partial}{\partial{\Delta_4}}-\frac{\partial}{\partial{\Delta_{2}}}}\Big)\widetilde M(z_i,\bz_i) = 0\ , \label{eq:A2_1} \\
&\Big( \partial_3-\frac{\Delta_3}{z_{23}}-\frac{1}{z_{13}} - \frac{\Delta_{2}-J_{2}-1+\bar{z}_{23}\bar{\partial}_{2}}{z_{23}} e^{\frac{\partial}{\partial{\Delta_3}}-\frac{\partial}{\partial{\Delta_{2}}}}\Big) \widetilde M(z_i,\bz_i)= 0\ , \label{eq:A2_2} \\
&\Big( \partial_3-\frac{\Delta_3}{z_{13}}-\frac{1}{z_{23}} - \frac{\Delta_{1}-J_{1}-1+\bar{z}_{13}\bar{\partial}_{1}}{z_{13}} e^{\frac{\partial}{\partial{\Delta_3}}-\frac{\partial}{\partial{\Delta_{1}}}}\Big) \widetilde M(z_i,\bz_i) = 0\ ,\label{eq:A2_3} \\
&\Big( \partial_4-\frac{\Delta_4}{z_{14}}-\frac{1}{z_{24}}- \frac{\Delta_{1}-J_{1}-1+\bar{z}_{14}\bar{\partial}_{1}}{z_{14}} e^{\frac{\partial}{\partial{\Delta_4}}-\frac{\partial}{\partial{\Delta_{1}}}}\Big)\widetilde M(z_i,\bz_i)= 0\ . \label{eq:A2_4}
\end{align}
Note that these are not standard partial differential equations because in addition to the  differential operators, they involve some parameter shifts induced by the operators $e^{\pm\frac{\partial}{\partial{\Delta}}}$ \cite{Stieberger1812}.
In order to take advantage of conformal symmetry, we define
\begin{align}
G(x,\bar{x}) &= \lim_{z_1,\bar{z}_1\rightarrow \infty} z_1^{2 h_1} \bar{z}_1^{2\bar{h}_1} M (z_1; z_2=1; z_3=x; z_4=0) \, , \\
\widetilde G(x,\bar{x}) &= \lim_{z_1,\bar{z}_1\rightarrow \infty} z_1^{2 h_1} \bar{z}_1^{2\bar{h}_1} \widetilde{M}(z_1; z_2=1; z_3=x; z_4=0)\, ,
\end{align}
with the conformal weights $h_1=(\Delta_1+J_1)/2, \bh_1=(\Delta_1-J_1)/2$. The complex variable $x$
is identified with the cross ratio
\be x= \frac{z_{12}z_{34}}{z_{13}z_{24}}\ .\ee
The partial amplitudes can be written as
\be
M(z_i,\bz_i)= \Pi(z_i,\bz_i)\,G(x,\bar{x})\ ,\qquad \widetilde M(z_i,\bz_i)= \Pi(z_i,\bz_i)\, \widetilde G(x,\bar{x})\ ,  \label{mmms}\ee
where
\begin{align}\Pi(z_i,\bz_i)&=
~z_{12}^{ ~ -h_1-h_2+h_3+h_4} z_{13}^{ ~-2h_3} z_{14}^{h_2+h_3-h_1-h_4}z_{24}^{h_1-h_2-h_3-h_4}  \nonumber\\
&\quad~\times \bar{z}_{12}^{ ~ -\bar{h}_1-\bar{h}_2+\bar{h}_3+\bar{h}_4} \bar{z}_{13}^{ ~-2\bar{h}_3} \bar{z}_{14}^{\bar{h}_2+\bar{h}_3-\bar{h}_1-\bar{h}_4}
\bar{z}_{24}^{\bar{h}_1-\bar{h}_2-\bar{h}_3-\bar{h}_4} .  \label{mmmss}
\end{align}
The dimensions of the primary fields associated with gluons (and in general, with all stable particles) are constrained by Re$(\Delta)=1$ \cite{Pasterski1705}. They are parameterized as $\Delta=1+i\lambda$, with real $\lambda$. After taking into account the helicities of external particles, the prefactor (\ref{mmmss}) becomes
\be
\Pi(z_i,\bz_i)= z_{12}^{~ -i \lambda_1- i\lambda_2+2} z_{13}^{~ -2-i\lambda_3}  z_{14}^{~ -i\lambda_1-i\lambda_4} z_{24}^{i\lambda_1-2}\bar{z}_{12}^{~ -i \lambda_1- i\lambda_2-2} \bar{z}_{13}^{~-i\lambda_3} \bar{z}_{14}^{~ -i\lambda_1-i\lambda_4}\bar{z}_{24}^{i\lambda_1}\ .
\ee

The BG equations (\ref{eq:A1_1}-\ref{eq:A2_4}) are equivalent to
\begin{align}
\Big[ 2-i \lambda_1 &-\frac{1+i\lambda_4}{x}+(x-1)\partial_x\Big] G(x,\bar{x}, \Delta_3,\Delta_4) \nonumber\\ &+\Big( \frac{i\lambda_3-1}{x}+\frac{\bar{x}\partial_{\bar{x}}}{x}\Big)G(x,\bar{x}, \Delta_3-1,\Delta_4+1) = 0\ , \label{eq:diff_G1_1} \\
&~ \nonumber\\
\Big( \partial_x ~+&~\frac{1+i\lambda_3}{x}-\frac{1}{1-x} \Big) G(x,\bar{x}, \Delta_3,\Delta_4) \nonumber\\
&+\Big[  \frac{1-i\lambda_4}{x} +\frac{\bar{x}}{x}\big( -i\lambda_1+(\bar{x}-1)\partial_{\bar{x}}\big)\Big]G(x,\bar{x}, \Delta_3+1,\Delta_4-1) = 0\ , \label{eq:diff_G1_2} \\
&~ \nonumber\\
\Big( \partial_x ~-&~\frac{1+i\lambda_3}{1-x}+\frac{1}{x}\Big) G(x,\bar{x},\Delta_3,\Delta_2) \nonumber\\
&-\Big[ \frac{1+i\lambda_2}{1-x}+\frac{1-\bar{x}}{1-x}(i\lambda_1-\bar{x}\partial_{\bar{x}})\Big]G(x,\bar{x},
\Delta_3+1,\Delta_2-1) = 0\ , \label{eq:diff_G1_3} \\
&~ \nonumber\\
 \Big( \partial_x ~+&~\frac{i\lambda_1-1}{1-x}+\frac{1}{x}\Big) G(x,\bar{x},\Delta_4,\Delta_1) \nonumber\\
&-\Big[ \frac{1-i\lambda_4}{1-x}+\frac{1-\bar{x}}{1-x}(-i\lambda_3-\bar{x}\partial_{\bar{x}})\Big] G(x,\bar{x},\Delta_4+1,\Delta_1-1) = 0\ , \label{eq:diff_G1_4}
\end{align}
and for the other group factor,
\begin{align}
\big[ 1-i\lambda_1-&i\lambda_4+(x-1)\partial_x\big] \widetilde G(x,\bar{x},\Delta_4,\Delta_2) \nonumber\\
&-\big(1+i\lambda_1+i\lambda_2-\bar{x}\partial_{\bar{x}}\big) \widetilde G(x,\bar{x},\Delta_4+1,\Delta_2-1) = 0\ , \label{eq:diff_G2_1} \\
&~ \nonumber\\
\Big( \partial_x -&\frac{1+i\lambda_3}{1-x}\Big)\widetilde  G(x,\bar{x},\Delta_3,\Delta_2)~~~~ \nonumber\\
&-\Big[ \frac{1+i\lambda_2}{1-x}+\frac{1-\bar{x}}{1-x}(i\lambda_1-\bar{x}\partial_{\bar{x}})\Big]
\widetilde G(x,\bar{x},\Delta_3+1,\Delta_2-1) = 0\ , \label{eq:diff_G2_2}
\end{align}
\begin{align}
\Big(\frac{1}{1-x}&-\partial_x\Big)\widetilde  G(x,\bar{x},\Delta_3,\Delta_1) \nonumber\\
+\big[ &(i\lambda_4+i\lambda_1-1)+(1-\bar{x})(i\lambda_3-i\lambda_1+\bar{x}\partial_{\bar{x}})\big] \widetilde G(x,\bar{x},\Delta_3+1,\Delta_1-1) = 0\ , \label{eq:diff_G2_3} \\
&~ \nonumber\\
 \Big( \partial_x +&\frac{i\lambda_1-1}{1-x}\Big) \widetilde G(x,\bar{x},\Delta_4,\Delta_1) \nonumber\\
&-\Big[ \frac{1-i\lambda_4}{1-x}+\frac{1-\bar{x}}{1-x}(-i\lambda_3-\bar{x}\partial_{\bar{x}})\Big] \widetilde G(x,\bar{x},\Delta_4+1,\Delta_1-1) = 0\ . \label{eq:diff_G2_4}
\end{align}
In the above equations, the shifts $\Delta_i\to\Delta_i\pm 1$ can be implemented by formal shifts
$i\lambda_i\to i\lambda_i\pm 1$.

Before diving into the details, one important comment is in order. As mentioned before, the Mellin transforms of four-gluon MHV amplitudes are zero, except on the real axis of $x=\bx$, where they are given by the distributions \cite{Strominger1706}:
\begin{align}
G(x,\bar{x})_{PSS} ~=~ &\delta(\lambda_1+\lambda_2+\lambda_3+\lambda_4)\,\label{pss1} x^{1-i\lambda_3-i\lambda_4}(1-x)^{-1+i\lambda_1+i\lambda_4}\delta(x-\bar{x})\ ,\\ \widetilde G(x,\bar{x})_{PSS} ~=~ & -x\,G(x,\bar{x})_{PSS}\ ,\label{pss2}
\end{align}
with $x\geq 1$ in the scattering channel $(12)\to (34)$ under consideration.
These amplitudes, however, do {\it not\/} satisfy Eqs.(\ref{eq:diff_G1_1})-(\ref{eq:diff_G2_4}). This is not surprising because they are not consistent with the OPEs of gluon operators following from the collinear limits of celestial amplitudes with a larger number of external particles, where the insertion points of the corresponding operators are allowed to move without violating the momentum conservation law
\cite{Fan1903}. This problem can possibly be alleviated by changing the metric signature to (2,2) and considering Klein spaces with the celestial torus replacing  celestial sphere at null infinity
\cite{Atanasov:2021oyu}. In this work, however, we do not entertain this possibility, remaining strictly in the framework of standard CFT \cite{DiF}.
\subsection{$s$-channel solutions $(x\approx 0)$}
Eqs.(\ref{eq:diff_G1_1})-(\ref{eq:diff_G2_4}) are separable in $x$ and $\bx$.
To examine the small $x$ behavior of $G(x,\bx)$, it is most convenient to start from Eqs.(\ref{eq:diff_G1_1}) and (\ref{eq:diff_G1_2}), and factorize the solutions as
\be
G_{n}(x,\bar{x},\Delta_3,\Delta_4) = f_n(x,\Delta_3,\Delta_4)~g_n(\bar{x},\Delta_3,\Delta_4) \, , \label{eq:G1nx}
\ee
where $n$ labels different separation constants $C_n$ and $D_n$ that will be introduced next. After inserting this form into Eq.(\ref{eq:diff_G1_1}), we obtain
\begin{align}
\frac{\big[~(2-i\lambda_1)x-(1+i\lambda_4)+x(x-1)\partial_x ~\big]f_n(x,\Delta_3,\Delta_4)}{f_n(x,\Delta_3-1,\Delta_4+1)} &= C_n\ ,\label{eq:fn1} \\[1mm]
\frac{\big[~(i\lambda_3-1)~+~\bar{x}\partial_{\bar{x}} ~\big]g_n(\bar{x},\Delta_3-1,\Delta_4+1)}{g_n(\bar{x},\Delta_3,\Delta_4)} =& -C_n\ . \label{eq:gn1}
\end{align}
Next, we shift $\Delta_3\rightarrow \Delta_3-1$, $\Delta_4\rightarrow \Delta_4+1$ in Eq.(\ref{eq:diff_G1_2}),
\begin{align}
\Big( x\partial_x ~+&~i\lambda_3-\frac{x}{1-x} \Big) G(x,\bar{x}, \Delta_3-1,\Delta_4+1) \nonumber\\
&+\Big[  -i\lambda_4 +\bar{x}\big( -i\lambda_1+(\bar{x}-1)\partial_{\bar{x}}\big)\Big]G(x,\bar{x}, \Delta_3,\Delta_4) = 0\ , \label{eq:diff_G1_2shift}
\end{align}
and separate it into
\begin{align}
\frac{\Big(x \partial_x +i\lambda_3+\frac{x}{1-x} \Big)f_n(x,\Delta_3-1,\Delta_4+1)}{f_n(x,\Delta_3,\Delta_4)} &= D_n \, , \label{eq:fn2} \\
 \frac{ \big[-i\lambda_4+\bar{x}(-i\lambda_1+(\bar{x}-1)\partial_{\bar{x}}) \big] g_n(\bar{x},\Delta_3,\Delta_4)}{g_n(\bar{x},\Delta_3-1,\Delta_4+1)} =& -D_n\ . \label{eq:gn2}
\end{align}
We combine Eqs.(\ref{eq:fn1}) and (\ref{eq:fn2}) to eliminate $f_n(x,\Delta_3-1,\Delta_4+1)$ and obtain an ordinary second order differential equation,
\begin{align}
\Big( x\frac{d}{dx} +i\lambda_3-\frac{x}{1-x} \Big)\Big[(2-i\lambda_1)x-(1+i\lambda_4)+&~x(x-1)\frac{d}{dx} \Big]f_n(x,\Delta_3,\Delta_4) \nonumber\\
&=E_n\ f_n(x,\Delta_3,\Delta_4)\ , \label{eq:fn3}
\end{align}
where $E_n=C_nD_n$.
Similarly, by combining (\ref{eq:gn1}) with (\ref{eq:gn2}), we obtain
\begin{align}
\Big(\bar{x}\frac{d}{d\bx}+i\lambda_3-1\Big)\Big[-i\lambda_4+\bar{x}\Big(-i\lambda_1+ &(\bar{x}-1)\frac{d}{d\bx}\Big) \Big] g_n(\bar{x},\Delta_3,\Delta_4) \nonumber\\[1mm]
=& ~E_n ~ g_n(\bar{x},\Delta_3,\Delta_4)  \, . \label{eq:gn3}
\end{align}

A priori, the separation constants $E_n$ are completely arbitrary. For any choice, Eqs.(\ref{eq:fn3}) and (\ref{eq:gn3}) can be transformed into  hypergeometric differential equations. Our choice of the separation constants will be determined by making contact with the single-valued amplitude derived in Ref.\cite{II}, where we have already observed that in the soft limit of $\lambda_1\to 0$, the amplitude satisfies BG equations. The result of Ref.\cite{II} can be written as
\be\lim_{\lambda_1=0}G(x,\bx)=\delta(\lambda_2+\lambda_3+\lambda_4)
\frac{(1+i\lambda_2)B(i\lambda_3,i\lambda_4)}{ x(1-x)}\ ,\label{lim1}\ee
modulo a multiplicative constant.\footnote{See Eq.(6.12) of Ref.\cite{II}, where we omitted the $\delta(\lambda_2+\lambda_3+\lambda_4)$ prefactor.} Also in this limit,
\be\widetilde G(x,\bx)=-x\, G(x,\bx)\ ,\label{lim2}\ee
as required by the symmetry of  full amplitude under exchanging the gluons $3\leftrightarrow 4$.
As we will see below, this limit is reached by the solution associated to the $n=0$ element of the following set:
\be
E_n = (n+i\lambda_4)(-n+1-i\lambda_3)\ ,
\ee
labeled by integer $n$. In the following discussion, the values of separation constants will be restricted to the above set.\footnote{A study of other choices turns out to lead to inconsistent factorization properties of CFT correlators.}

Eqs.(\ref{eq:fn3}) and (\ref{eq:gn3}) can be transformed into hypergeometric equations. Each of them has two solutions: $\big\{f^{(a)}_n $, $f^{(b)}_n \}$ and $\{g^{(a)}_n $, $g^{(b)}_n \}$, respectively.
Only two products, however, are single-valued near $x=0$:
\begin{align}
a_n\, f_{n}^{(a)}g_{n}^{(a)} =& ~\frac{a_n}{x(1-x)}~ x^n ~ {}_2F_1\left({-i\lambda_1+n,\, i\lambda_3+n\atop i\lambda_3+i\lambda_4+2n};\,x\right)  \bar{x}^n ~ {}_2F_1\left({-i\lambda_1+n,\, i\lambda_3+n\atop i\lambda_3+i\lambda_4+2n};\,\bar{x}\right)\ , \label{eq:fngn1} \\
b_n\, f_{n}^{(b)}g_{n}^{(b)} =& ~\frac{b_n}{x(1-x)} ~x^{~-n+1-i\lambda_3-i\lambda_4}~ {}_2F_1\left({-n+1 -i\lambda_4, \, -n+1+i\lambda_2\atop 2-2n-i\lambda_3-i\lambda_4};\,x\right) \nonumber\\[1mm]
&~~~~~~~~~~\times \bar{x}^{~-n+1-i\lambda_3-i\lambda_4}~ {}_2F_1\left({-n+1 -i\lambda_4,\, -n+1+i\lambda_2\atop 2-2n-i\lambda_3-i\lambda_4};\,\bar{x}\right)\ , \label{eq:fngn2}
\end{align}
where $a_n$ and $b_n$ are coefficients that are arbitrary at this point, but will be determined later. Note that $a_0$ is the coefficient of a term that behaves as $x^{-1}$ near $x=0$, therefore it must be related to the $\lambda_1\to 0$ limit written in Eq.(\ref{lim1}). We clearly need the $n=0$ solution to make contact with this limit. Once we insert it, however, into Eqs.(\ref{eq:diff_G1_3}) and (\ref{eq:diff_G1_4}), we find that they are not satisfied. The remainder is of order $x^{-1}\bx^1$, therefore we can get closer to the solution by adding $a_1\, f_{1}^{(a)}g_{1}^{(a)}$.
Recursively, we need all $a_n\, f_{n}^{(a)} g_{n}^{(a)}$ with $n\geq 0$. As a result, the first solution of Eqs.(\ref{eq:diff_G1_1})-(\ref{eq:diff_G1_4}) takes the form:
\begin{align}
G(x,\bar{x})_a &= \sum_{n=0}^{\infty} a_nf_{n}^{(a)} g_{n}^{(a)} \nonumber\\
&=\sum_{n=0}^{\infty} a_n \frac{1}{x(1-x)}\,x^n {}_2F_1\left({-i\lambda_1+n,\, i\lambda_3+n\atop i\lambda_3+i\lambda_4+2n};\,x\right)  \bar{x}^n {}_2F_1\left({-i\lambda_1+n,\, i\lambda_3+n\atop i\lambda_3+i\lambda_4+2n};\,\bar{x}\right) \, . \label{eq:sols_G1a}
\end{align}
The coefficients $a_n$ can be determined recursively by plugging the above ansatz back into Eqs.(\ref{eq:diff_G1_1})-(\ref{eq:diff_G1_4}) and using Gau\ss\ contiguous relations for hypergeometric functions.  For $a_0$, we obtain
\begin{align}
-i\lambda_4 ~a_0(\Delta_3,\Delta_4)& +(i\lambda_3-1)~a_0(\Delta_3-1,\Delta_4+1) =0 \, ,\\[1mm]
i\lambda_3 ~ a_0(\Delta_3,\Delta_4) & +(1-i\lambda_4) ~ a_0(\Delta_3+1,\Delta_4-1) =0 \, , \\
 \frac{i\lambda_2~ i\lambda_3}{i\lambda_3+i\lambda_4} ~a_0(\Delta_3,\Delta_2) & -(1+i\lambda_1+i\lambda_2) ~ a_0(\Delta_3+1,\Delta_2-1) =0 \, ,\\
 \frac{i\lambda_1~ i\lambda_4}{i\lambda_3+i\lambda_4} ~a_0(\Delta_4,\Delta_1) & - (1+i\lambda_1+i\lambda_2) ~ a_0(\Delta_4+1,\Delta_1-1) =0 \, .
\end{align}
These relations are solved by
\be
a_0 =\delta(\lambda_1+\lambda_2+\lambda_3+\lambda_4)
 (1+i\lambda_1+i\lambda_2)B(-i\lambda_1,-i\lambda_2)B(i\lambda_3,i\lambda_4)\ .\label{azsol}
\ee
Note that in the $\lambda_1\to 0$ limit,
\be\lim_{\lambda_1=0}i\lambda_1a_0=\delta(\lambda_2+\lambda_3+\lambda_4)
(1+i\lambda_2)B(i\lambda_3,i\lambda_4)\ ,\label{lim3}\ee
in agreement with Eq.(\ref{lim1}) modulo a (infinite) numerical factor.

In the second step of recursive procedure,
Eqs.(\ref{eq:diff_G1_1})-(\ref{eq:diff_G1_4}) yield the following relations between $a_0$ and $a_1$:
\begin{align}
(-i\lambda_4-1)~a_1(\Delta_3,\Delta_4) + i\lambda_3 ~a_1(\Delta_3& -1,\Delta_4+1) =0 \, , \\[1mm]
(i\lambda_3+1) ~a_1(\Delta_3,\Delta_4)  -i\lambda_4 ~a_1(\Delta_3&+1,\Delta_4-1) = 0 \, , \\
 a_0(\Delta_3+1,\Delta_2-1) \frac{i\lambda_1~i\lambda_4 ~(i\lambda_3+1)(i\lambda_2-1)}{(i\lambda_3+i\lambda_4+1)^2 (i\lambda_3+i\lambda_4+2)} &+(i\lambda_1+i\lambda_2) a_1(\Delta_3+1,\Delta_2-1) = 0 \, , \\
 a_0(\Delta_4+1,\Delta_1-1) \frac{i\lambda_2~i\lambda_3 ~(i\lambda_4+1)(i\lambda_1-1)}{(i\lambda_3+i\lambda_4+1)^2 (i\lambda_3+i\lambda_4+2)} &+(i\lambda_1+i\lambda_2) a_1(\Delta_4+1,\Delta_1-1) = 0 \, ,
\end{align}
which are solved by
\be
a_1=\delta(\lambda_1+\lambda_2+\lambda_3+\lambda_4)
(i\lambda_1+i\lambda_2-1)B(1-i\lambda_1,1-i\lambda_2)B(1+i\lambda_3,1+i\lambda_4) \, .
\ee
After repeating the same steps for $a_n$, we find
\be
a_n= \delta(\lambda_1+\lambda_2+\lambda_3+\lambda_4)
(1+i\lambda_1+i\lambda_2-2n)B(n-i\lambda_1,n-i\lambda_2)B(n+i\lambda_3,n+i\lambda_4) \, .
\ee
In this way, we complete the construction of the first solution,
$G(x,\bar{x})_a$, written as a series in $x$ and $\bar{x}$ in Eq.(\ref{eq:sols_G1a}), with the  coefficients given above. Note that in the $\lambda_1\to 0$ limit, all coefficients $a_n$ with $n>0$ are finite, as contrasted with infinite $a_0$; therefore, $G(x,\bar{x})_a$ does indeed reproduce the amplitude obtained in Ref.\cite{II}.

Looking back at Eq.(\ref{eq:fngn1}), we could have also started from $a_{-1}f_{-1}^{(a)}\,g_{-1}^{(a)} $ and tried to build up another solution. But once we plug it into Eqs.(\ref{eq:diff_G1_3}) and (\ref{eq:diff_G1_4}), we find that we need to introduce $a_{-2}f_{-2}^{(a)}\,g_{-2}^{(a)} $ and we end up with an infinite series of negative powers of $x$ and $\bar{x}$. This series diverges for small $|x|$, therefore we reject it on physical grounds. In the case of the second solution, see Eq.(\ref{eq:fngn2}), such an unphysical series appears when we start from $b_{0}f_{0}^{(b)}\,g_{0}^{(b)}$, therefore a physically acceptable solution can possibly be built  by starting from  $b_{-1}f_{-1}^{(b)}\,g_{-1}^{(b)}$ only. It takes the  form:
\begin{align}
G(x,\bar{x})_b &= \sum_{n=-1}^{-\infty} b_n ~f_{n}^{(b)}~ g_{n}^{(b)}= \sum_{m=0}^{+\infty} b_{m} ~f_{-m-1}^{(b)}~ g_{-m-1}^{(b)} \nonumber\\
&=\sum_{m=0}^{+\infty} b_m  \frac{1}{x(1-x)} ~x^{2-i\lambda_3-i\lambda_4+m}~ {}_2F_1\left({2 -i\lambda_4+m, \, 2+i\lambda_2+m\atop 4-i\lambda_3-i\lambda_4+2m};\,x\right) \nonumber\\
&~~~~~~\times \bar{x}^{2-i\lambda_3-i\lambda_4+m}~ {}_2F_1\left({2 -i\lambda_4+m,\, 2+i\lambda_2+m\atop 4-i\lambda_3-i\lambda_4+2m};\,\bar{x}\right) \, , \label{eq:sols_G1b}
\end{align}
where in the first line, we relabeled the index $n$ as $m=-n-1$. The coefficients $b_m$ can be determined in the same way as $a_n$, with the following result:
\begin{align}
b_m = &~\delta(\lambda_1+\lambda_2+\lambda_3+\lambda_4)
(i\lambda_3+i\lambda_4-3-2m)\nonumber\\ &\times B(m+2+i\lambda_1,m+2+i\lambda_2)B(m+2-i\lambda_3,m+2-i\lambda_4) \, .
\end{align}
Note that all these coefficients are finite in the $\lambda_1\to 0$ limit.
While the normalization of the first solution $G(x,\bar{x})_a$ is determined by the $\lambda_1\to 0$ limit, the overall normalization of the second solution  $G(x,\bar{x})_b$ remains arbitrary at this point.  In this way, we obtain
\be G(x,\bx)= G(x,\bx)_a+\alpha G(x,\bx)_b\ .\label{ssol}\ee
The coefficient $\alpha$ will be fixed later to $\alpha=-1$ by gluing it to the solutions valid in the neighborhood of $x= 1$.

In order to solve Eqs.(\ref{eq:diff_G2_1})-(\ref{eq:diff_G2_4}) associated with the second group factor, we start from the observation that when $\widetilde G=-x\,G$, {\it i.e}.\ the relation that holds in the $\lambda_1=0$ limit, is inserted into Eq.(\ref{eq:diff_G2_2}), it takes the form of
Eq.(\ref{eq:diff_G1_3}), which is satisfied by the functions written in Eqs.(\ref{eq:sols_G1a}) and (\ref{eq:sols_G1b}). This leads to the proposition that
\begin{align}\label{ssol1}
\widetilde G(x,\bar{x})_a= -x\, G(x,\bar{x})_a\ , \qquad \widetilde G(x,\bar{x})_b= -x\,G(x,\bar{x})_b\ .
\end{align}
Indeed, both functions satisfy Eqs.(\ref{eq:diff_G2_1})-(\ref{eq:diff_G2_4}).

\subsection{$u$-channel solutions $(x\approx 1)$}
Since BG equations are separable, with the solutions given by the products of holomorphic and antiholomorphic functions, one could try to extend them from the region of small $|x|$ to the entire complex plane by analytically continuing the holomorphic and antiholomorphic factors. Unfortunately, in the neighborhood of $x=0$, our solutions are given by infinite series, therefore such a na\"ive procedure is doomed to fail due to convergence problems. Instead, we will start from scratch, by constructing the solutions valid near $x=1$ and separately, at $x\to\infty$, and only at the end, we will glue all solutions together into one single-valued function. In the region $x\approx 1$,
it is convenient to start from Eq.(\ref{eq:diff_G1_3}) and (\ref{eq:diff_G1_4}). They are solved by
\begin{align}
c_n\, r_{n}^{(c)}(1-x)\,t_{n}^{(c)}(1-\bar{x}) = \frac{c_n}{x(1-x)}& (1-x)^n \, _2F_1\left({n-i\lambda_1, n+i\lambda_3 \atop 2n+2-i\lambda_1-i\lambda_4}; 1-x\right)    \nonumber\\
&\times (1-\bar{x})^n \, _2F_1\left({n-i\lambda_1, n+i\lambda_3 \atop 2n+2-i\lambda_1-i\lambda_4}; 1-\bar{x}\right) \, , \end{align}\begin{align}
d_n\,r_{n}^{(d)}(1-x)\,t_{n}^{(d)}&(1-\bar{x})\nonumber=\\ =& \,\frac{d_n}{x(1-x)} (1-x)^{-n-1+i\lambda_1+i\lambda_4} \, _2F_1\left( {-n-1-i\lambda_2,-n-1+i\lambda_4\atop -2n +i\lambda_1+i\lambda_4};1-x\right) \nonumber\\
&\times (1-\bar{x})^{-n-1+i\lambda_1+i\lambda_4} \, _2F_1\left( {-n-1-i\lambda_2,-n-1+i\lambda_4\atop -2n +i\lambda_1+i\lambda_4};1-\bar{x}\right) \, ,
\end{align}
where the coefficients $c_n$ and $d_n$ remain  to be determined.
{}From here, we proceed in the same way as  after Eq.(\ref{eq:fngn2}). Starting from
$c_0 r_{0}^{(c)}t_{0}^{(c)}$,  we can generate a solution of  Eqs.(\ref{eq:diff_G1_1})-(\ref{eq:diff_G1_2}) by adding an infinite series of $c_n r_{n}^{(c)}t_{n}^{(c)}$  with increasing powers  of $(1-x)$, {\it i.e}.\ $n\geq0$. For $c_0$, we obtain
\begin{align}
&i\lambda_3\, c_0(\Delta_3,\Delta_2)+(1+i\lambda_2)c_0(\Delta_3+1,\Delta_2-1) = 0 \, ,\\[1mm]
&i\lambda_1 \, c_0(\Delta_4,\Delta_1) -(1-i\lambda_4)c_0(\Delta_4+1,\Delta_1-1) = 0 \, , \\
&(i\lambda_2+i\lambda_3)c_0(\Delta_3,\Delta_4) +\frac{(i\lambda_3-1)(1-i\lambda_4)}{1-i\lambda_1-i\lambda_4}c_0(\Delta_3-1,\Delta_4+1) = 0 \, , \\
& \frac{i\lambda_3 (2-i\lambda_4)}{2-i\lambda_1-i\lambda_4}c_0(\Delta_3,\Delta_4) +(1+i\lambda_2+i\lambda_3)c_0(\Delta_3+1,\Delta_4-1) = 0 \, .
\end{align}
In this case, we find two solutions:
\begin{align}
c_{0}^{(1)} &= \delta(\lambda_1+\lambda_2+\lambda_3+\lambda_4) B(i\lambda_3,-i\lambda_2-i\lambda_3-1)B(-i\lambda_1,i\lambda_4+i\lambda_1-1)(1-i\lambda_1-i\lambda_4) \, ,\label{c01}\\[1mm]
c_{0}^{(2)} &= \delta(\lambda_1+\lambda_2+\lambda_3+\lambda_4) B(2+i\lambda_2,-i\lambda_2-i\lambda_3-1)B(2-i\lambda_4,i\lambda_4+i\lambda_1-1)(1+i\lambda_2+i\lambda_3) \, .
\end{align}
From here, we can recursively determine two sets, $c_{n}^{(1)}$ and $c_{n}^{(2)}$ $(n\geq 0)$:
\begin{align}
c_{n}^{(1)} &=\delta(\lambda_1+\lambda_2+\lambda_3+\lambda_4)(1-i\lambda_1-i\lambda_4+2n)\nonumber \\& ~~~~~\times B(i\lambda_3+n,-i\lambda_2-i\lambda_3-1-2n) B(-i\lambda_1+n,i\lambda_4+i\lambda_1-1-2n)\label{c11}
\, ,\\[1mm]
c_{n}^{(2)} &= \delta(\lambda_1+\lambda_2+\lambda_3+\lambda_4)(1+i\lambda_2+i\lambda_3+2n) \nonumber\\ & ~~~~~\times B(2+i\lambda_2+n,-i\lambda_2-i\lambda_3-1-2n) B(2-i\lambda_4+n,i\lambda_4+i\lambda_1-1-2n)
\, .
\end{align}
Next, we start from $d_{-1}\,r_{-1}^{(d)}\,t_{-1}^{(d)}$, and after repeating similar steps, we find that
we can  construct the solutions of  Eqs.(\ref{eq:diff_G1_1})-(\ref{eq:diff_G1_2}) by adding
$d_{n}\,r_{n}^{(d)}\,t_{n}^{(d)}$ with $n\leq-1$, to form two distinct series with
increasing powers of $(1-x)$. Here again, the coefficients can be determined recursively.
After relabeling the indices $-n-1\to n\ge 0$, we obtain
\begin{align}
d_{n}^{(1)} & =\delta(\lambda_1+\lambda_2+\lambda_3+\lambda_4)(1-i\lambda_2-i\lambda_3+2n)\nonumber \\&
~~~~~\times B(i\lambda_4+n,-i\lambda_1-i\lambda_4-1-2n)B(n-i\lambda_2,i\lambda_3+i\lambda_2-1-2n)
\, ,\\[1mm]
d_{n}^{(2)} &=\delta(\lambda_1+\lambda_2+\lambda_3+\lambda_4)(1+i\lambda_1+i\lambda_4+2n) \nonumber\\ &
~~~~~\times B(n+2+i\lambda_1,-i\lambda_1-i\lambda_4-1-2n)B(n+2-i\lambda_3,i\lambda_3+i\lambda_2-1-2n)\label{d11}
\, .
\end{align}

As a result, we obtain four solutions of BG equations valid in the region of $x\approx 1$:
\begin{align}
G(1-x,1-\bar{x})_a
&= \frac{1}{x(1-x)} \sum_{n=0}^{+\infty} c_n^{(1)} (1-x)^n \, _2F_1\left({n-i\lambda_1, n+i\lambda_3 \atop 2n+2-i\lambda_1-i\lambda_4}; 1-x\right)\nonumber\\
&~~~~~~~\times (1-\bar{x})^n \, _2F_1\left({n-i\lambda_1, n+i\lambda_3 \atop 2n+2-i\lambda_1-i\lambda_4}; 1-\bar{x}\right) \, , \label{eq:solu_G1a} \\
G(1-x,1-\bar{x})_b
&=\frac{1}{x(1-x)} \sum_{n=0}^{+\infty} c_n^{(2)}(1-x)^n \, _2F_1\left({n-i\lambda_1, n+i\lambda_3 \atop 2n+2-i\lambda_1-i\lambda_4}; 1-x\right) \nonumber\\
&~~~~~~~\times (1-\bar{x})^n \, _2F_1\left({n-i\lambda_1, n+i\lambda_3 \atop 2n+2-i\lambda_1-i\lambda_4}; 1-\bar{x}\right) \, , \label{eq:solu_G1b}
\end{align}
\begin{align}
G(1-x,1-\bar{x})_c
&=\frac{1}{x(1-x)} \sum_{n=0}^{+\infty}
d_n^{(1)} (1-x)^{n+i\lambda_1+i\lambda_4}\, _2F_1\left({n-i\lambda_2, n+i\lambda_4 \atop 2n+2+i\lambda_1+i\lambda_4};1-x\right)\nonumber\\
&~~~~~~~\times (1-\bar{x})^{n+i\lambda_1+i\lambda_4}\, _2F_1\left({n-i\lambda_2, n+i\lambda_4 \atop 2n+2+i\lambda_1+i\lambda_4};1-\bar{x}\right) \, , \label{eq:solu_G1c}\\
G(1-x,1-\bar{x})_d
&=\frac{1}{x(1-x)} \sum_{n=0}^{+\infty} d_n^{(2)} (1-x)^{n+i\lambda_1+i\lambda_4}\, _2F_1\left({n-i\lambda_2, n+i\lambda_4 \atop 2n+2+i\lambda_1+i\lambda_4};1-x\right)\nonumber\\
&~~~~~~~\times (1-\bar{x})^{n+i\lambda_1+i\lambda_4}\, _2F_1\left({n-i\lambda_2, n+i\lambda_4 \atop 2n+2+i\lambda_1+i\lambda_4};1-\bar{x}\right) \, , \label{eq:solu_G1d}
\end{align}
with the coefficients given in Eqs.(\ref{c11})-(\ref{d11}).

In the $\lambda_1\to 0$ limit, the coefficient $c_0^{(1)}$ (\ref{c01}) blows up,
\be\lim_{\lambda_1=0}i\lambda_1c_0^{(1)}=\delta(\lambda_2+\lambda_3+\lambda_4)
(1+i\lambda_2)B(i\lambda_3,i\lambda_4)\ ,\label{lim4}\ee
while all other coefficients are finite. Hence the normalization of $G(1-x,1-\bar{x})_a$ is fixed by the $\lambda_1\to 0$ limit of the single-valued amplitude \cite{II}.
The relative factor between \linebreak $G(1-x,1-\bar{x})_d $ and $G(1-x,1-\bar{x})_a$ is determined by the cancellation of an unphysical singularity of the series coefficients at $\lambda_2+\lambda_3=0$. The relative factor between $G_1(1-x,1-\bar{x})_b$ and $G_1(1-x,1-\bar{x})_c$ is also determined in a similar way. Hence, the only undetermined factor is the normalization of $G_1(1-x,1-\bar{x})_b$. Actually, it can be fixed
by matching with
the known OPEs, discussed in Section 4. In this way, we obtain
\be
G(1-x,1-\bar{x})= G(1-x,1-\bar{x})_a - G(1-x,1-\bar{x})_b+G(1-x,1-\bar{x})_c-G(1-x,1-\bar{x})_d \, .\label{eq:s=u}
\ee

\subsection{$t$-channel solutions $(x\to \infty)$}
In order to find solutions valid at large $|x|$, in the form of series expansions in powers of $1/x$,  it is convenient to start
from Eqs.(\ref{eq:diff_G2_1}) and  (\ref{eq:diff_G2_3}). Then the solution to the full set of BG equations can be constructed in a similar way as for small $x$ and $x\approx 1$. We obtain four solutions:
\begin{align}
G\left(\frac{1}{x},\frac{1}{\bar{x}}\right)_a& =  \frac{1}{x(1-x)} \sum_{n=0}^{+\infty}
 t^{(1)}_n
 \left(\frac{1}{x}\right)^{n-i\lambda_1} \, _2F_1\left({n+2+i\lambda_2,n-i\lambda_1 \atop 2-i\lambda_3-i\lambda_1+2n}; \frac{1}{x}\right)\nonumber\\ &~~~~~~~~\times\left(\frac{1}{\bar{x}}\right)^{n-i\lambda_1} \, _2F_1\left({n+2+i\lambda_2,n-i\lambda_1 \atop 2-i\lambda_3-i\lambda_1+2n}; \frac{1}{\bar{x}}\right) \, , \label{eq:solt_G1a} \\[1mm]
G\left(\frac{1}{x},\frac{1}{\bar{x}}\right)_b &=
\frac{1}{x(1-x)} \sum_{n=0}^{+\infty} t^{(2)}_n  \left(\frac{1}{x}\right)^{n-i\lambda_1} \, _2F_1\left({n+2+i\lambda_2,n-i\lambda_1 \atop 2-i\lambda_3-i\lambda_1+2n}; \frac{1}{x}\right)\nonumber\\ &~~~~~~~~\times\left(\frac{1}{\bar{x}}\right)^{n-i\lambda_1} \, _2F_1\left({n+2+i\lambda_2,n-i\lambda_1 \atop 2-i\lambda_3-i\lambda_1+2n}; \frac{1}{\bar{x}}\right) \, , \label{eq:solt_G1b}
\end{align}
\begin{align}
G\left(\frac{1}{x},\frac{1}{\bar{x}}\right)_c &=
 \frac{1}{x(1-x)} \sum_{n=0}^{+\infty} u^{(1)}_n   \left(\frac{1}{x}\right)^{n+i\lambda_3}\, _2F_1\left( {2-i\lambda_4+n, i\lambda_3+n\atop 2+i\lambda_1+i\lambda_3+2n}; \frac{1}{x}\right)\nonumber\\ &~~~~~~~~\times\left(\frac{1}{\bar{x}}\right)^{n+i\lambda_3}\, _2F_1\left( {2-i\lambda_4+n, i\lambda_3+n\atop 2+i\lambda_1+i\lambda_3+2n}; \frac{1}{\bar{x}}\right) \, ,    \label{eq:solt_G1c} \\
G\left(\frac{1}{x},\frac{1}{\bar{x}}\right)_d &=
\frac{1}{x(1-x)} \sum_{n=0}^{+\infty} u^{(2)}_n
\left(\frac{1}{x}\right)^{n+i\lambda_3}\, _2F_1\left( {2-i\lambda_4+n, i\lambda_3+n\atop 2+i\lambda_1+i\lambda_3+2n}; \frac{1}{x}\right)\nonumber\\ &~~~~~~~~\times\left(\frac{1}{\bar{x}}\right)^{n+i\lambda_3}\, _2F_1\left( {2-i\lambda_4+n, i\lambda_3+n\atop 2+i\lambda_1+i\lambda_3+2n}; \frac{1}{\bar{x}}\right) \, . \label{eq:solt_G1d}
\end{align}
with the coefficients
\begin{align}
t^{(1)}_n&=\delta(\lambda_1+\lambda_2+\lambda_3+\lambda_4)(1-i\lambda_1-i\lambda_3+2n) \nonumber\\ &~~~~~\times B(i\lambda_4+n,-i\lambda_2-i\lambda_4-1-2n)B(n-i\lambda_1,i\lambda_3+i\lambda_1-1-2n)\ ,\\
t^{(2)}_n&=\delta(\lambda_1+\lambda_2+\lambda_3+\lambda_4)(1+i\lambda_2+i\lambda_4+2n) \nonumber\\
&~~~~~\times B(2+i\lambda_2+n,-i\lambda_2-i\lambda_4-1-2n)B(2-i\lambda_3+n,i\lambda_3+i\lambda_1-1-2n) \ ,\\
u^{(1)}_n&=\delta(\lambda_1+\lambda_2+\lambda_3+\lambda_4)(1-i\lambda_2-i\lambda_4+2n)\nonumber\\ &~~~~~\times B(i\lambda_3+n,-i\lambda_1-i\lambda_3-1-2n)B(n-i\lambda_2,i\lambda_2+i\lambda_4-1-2n)     \ ,\\
u^{(2)}_n&=\delta(\lambda_1+\lambda_2+\lambda_3+\lambda_4)(1+i\lambda_1+i\lambda_3+2n)\nonumber\\ &~~~~~\times B(2+i\lambda_1+n,-i\lambda_1-i\lambda_3-1-2n)B(2-i\lambda_4+n,i\lambda_2+i\lambda_4-1-2n)\ .
\end{align}
In this case, only the coefficient $t^{(1)}_0$ blows up in the
$\lambda_1\to 0$ limit,
\be\lim_{\lambda_1=0}i\lambda_1t_0^{(1)}=\delta(\lambda_2+\lambda_3+\lambda_4)
(1+i\lambda_2)B(i\lambda_3,i\lambda_4)\ ,\label{lim5}\ee
which determines the normalization of $G\left(\frac{1}{x},\frac{1}{\bar{x}}\right)_a$. The relative normalization of the other three series can be determined in a similar way as for $x\approx 1$.
In this way, we obtain
\be  G\left(\frac{1}{x},\frac{1}{\bar{x}}\right)=
 G\left(\frac{1}{x},\frac{1}{\bar{x}}\right)_a -G\left(\frac{1}{x},\frac{1}{\bar{x}}\right)_b+G\left(\frac{1}{x},\frac{1}{\bar{x}}\right)_c-
 G\left(\frac{1}{x},\frac{1}{\bar{x}}\right)_d \, .\label{tsol}\ee
\subsection{Single-valued four-gluon amplitude}
At this point, we have three solutions of BG equations, Eqs.(\ref{ssol}), (\ref{eq:s=u}) and (\ref{tsol}), written as series expansions near $x= 0$,  $x=1$ and $x\to \infty$, respectively.
They should be equal to the expansions of one single-valued solution in the respective regions. This is not easy to check analytically because the series do not sum to closed-form expressions. All that we can do is to compare the expansions numerically in the overlapping regions. In particular, both $u$-channel and $t$-channel expansions should be valid at some points ``between'' 1 and infinity. We studied
Eqs.(\ref{eq:s=u}) and (\ref{tsol}) near $x=1.5$, for general complex arguments, and found that with
$10^{-10}$ accuracy
\begin{align}
 G(1-x&,1-\bar{x})_a - G(1-x,1-\bar{x})_b+G(1-x,1-\bar{x})_c-G(1-x,1-\bar{x})_d \nonumber\\
&= ~G\left(\frac{1}{x},\frac{1}{\bar{x}}\right)_a -G\left(\frac{1}{x},\frac{1}{\bar{x}}\right)_b+G\left(\frac{1}{x},\frac{1}{\bar{x}}\right)_c-
G\left(\frac{1}{x},\frac{1}{\bar{x}}\right)_d \, , \label{eq:u=t}
\end{align}
We also compared $s$- and $u$-channel expansions near $x=0.5$. We found agreement with $10^{-10}$ accuracy, provided that  the coefficient $\alpha=-1$ in Eq.(\ref{ssol}):
\begin{align} &G(x,\bar{x})_a  - G(x,\bar{x})_b\nonumber\\
 &= G(1-x,1-\bar{x})_a - G(1-x,1-\bar{x})_b+G(1-x,1-\bar{x})_c-G(1-x,1-\bar{x})_d \nonumber\\
\end{align}
We conclude that Eqs.(\ref{ssol}), (\ref{eq:s=u}) and (\ref{tsol}) describe one single-valued function.
In this way, we obtain the following single-valued four-gluon CCFT correlator:
\begin{align}
\Big\langle\phi_{\D_1,-}^{a_1,-\epsilon}&(z_{1},\zbar_{1}
)\,
\phi_{\D_2,-}^{a_2,-\epsilon}(z_2,\bz_2)
\,\phi_{\D_3,+}^{a_3,+\epsilon}
(z_3,\bar z_3)
\,\phi_{\D_4,+}^{a_4,+\epsilon}(z_4,\bz_4)\Big\rangle_{\text{SV}} \nonumber\\[1mm]
&=  z_{12}^{~ -i \lambda_1- i\lambda_2+2} z_{24}^{i\lambda_1-2} z_{14}^{~ -i\lambda_1-i\lambda_4} z_{13}^{~ -2-i\lambda_3} \bar{z}_{12}^{~ -i \lambda_1- i\lambda_2-2} \bar{z}_{24}^{i\lambda_1} \bar{z}_{14}^{~ -i\lambda_1-i\lambda_4} \bar{z}_{13}^{~-i\lambda_3} \nonumber\\[2mm]
&~~~~~~~\times \big[f^{a_1a_2b}f^{a_3a_4b}G_{\text{SV}}(x,\bx)+f^{a_1a_3b}f^{a_2a_4b} \widetilde G_{\text{SV}}(x,\bx)\big]\, , \label{eq:full_MHV_SV}
\end{align}
with
\begin{align}
&G_{\text{SV}}(x,\bx) \nonumber\\
&=
\begin{cases}
G(x,\bar{x})_a-G(x,\bar{x})_b & (x\approx 0)\\
G(1-x,1-\bar{x})_a - G(1-x,1-\bar{x})_b+G(1-x,1-\bar{x})_c-G(1-x,1-\bar{x})_d & (x\approx 1) \\
G\left(\frac{1}{x},\frac{1}{\bar{x}}\right)_a -G\left(\frac{1}{x},\frac{1}{\bar{x}}\right)_b+G\left(\frac{1}{x},\frac{1}{\bar{x}}\right)_c-
G\left(\frac{1}{x},\frac{1}{\bar{x}}\right)_d &\!\! (x\!\to\!\infty)~~~~
\end{cases} \label{eq:sol_G1_sv}
\end{align}and
\be
\widetilde G_{\text{SV}}(x,\bx) = -x\, G_{\text{SV}}(x,\bx) \, . \label{eq:sol_G2_sv}
\ee
\subsection{Single-valued three-gluon amplitude}
We can extract the three-gluon MHV correlator by taking the $z_4\to z_3$ limit of the correlator
(\ref{eq:full_MHV_SV}) and using the leading term in the OPE of two gluon fields\footnote{Recall that $\Delta=1+i\lambda$ for the primary fields associated with gluons.} \cite{Fan1903,Strominger1910}:
\begin{align}
\phi_{\D_3,+}^{a_3,+\epsilon}(z_3,\bar z_3)\,\phi_{\D_4,+}^{a_4,+\epsilon}(z_4,\bz_4) =
\frac{-i f^{a_3 a_4 x}}{z_{34}} B(i\lambda_3,i\lambda_4)\phi_{\Delta=1+i\lambda_3+i\lambda_4,+}^{x,+\epsilon}(z_3,\bz_3) \ , \label{eq:z34OPE1}\end{align}
In this limit, $x\to 0$, and the dominant contribution to the four-gluon correlator comes from the $n=0$
term in $G(x,\bar{x})_a$, see Eqs.(\ref{eq:sols_G1a}) and (\ref{azsol}). We have
\begin{align}
\Big\langle\phi_{\D_1,-}^{a_1,-\epsilon}&(z_{1},\zbar_{1}
)\,
\phi_{\D_2,-}^{a_2,-\epsilon}(z_2,\bz_2)
\,\phi_{\D_3,+}^{a_3,+\epsilon}
(z_3,\bar z_3)
\,\phi_{\D_4,+}^{a_4,+\epsilon}(z_4,\bz_4)\Big\rangle_{\text{SV}} \nonumber\\
&\approx f^{a_1a_2x}f^{a_3a_4x}\delta(\lambda_1+\lambda_2+\lambda_3+\lambda_4)\, (1+i\lambda_1+\lambda_2)B(-i\lambda_1,-i\lambda_2)B(i\lambda_3,i\lambda_4)\nonumber\\[1mm]
&~~~~~~~~~~~~~~~~~~\times z_{34}^{-1}\, z_{12}^{\, 1-i\lambda_1-i\lambda_2} z_{23}^{\, i\lambda_1-1}z_{13}^{\, i\lambda_2-1} \bz_{12}^{\, -i\lambda_1-i\lambda_2-2} \bz_{23}^{\, i\lambda_1} \bz_{13}^{\, i\lambda_2}\ .
\label{eq:z34limit}
\end{align}
By comparing with Eq.(\ref{eq:z34OPE1}), we find
\begin{align}
\Big\langle \phi_{\D_1,-}^{a_1,-\epsilon}&(z_{1},\zbar_{1})\, \phi_{\D_2,-}^{a_2,-\epsilon}(z_2,\bz_2)\,\phi_{\D_3,+}^{a_3,+\epsilon}(z_3,\bz_3)
\Big\rangle_{\text{SV}} \nonumber\\
&= \, i f^{a_1a_2 a_3}\, \delta(\lambda_1+\lambda_2+\lambda_3)\, (1+i\lambda_1+\lambda_2)B(-i\lambda_1,-i\lambda_2) \nonumber\\[1mm]
&~~~~~~~~~~\times z_{12}^{\, 1-i\lambda_1-i\lambda_2} \bz_{12}^{\, -i\lambda_1-i\lambda_2-2}\,z_{23}^{\, i\lambda_1-1}\bz_{23}^{\, i\lambda_1}\,z_{13}^{\, i\lambda_2-1}  \bz_{13}^{\, i\lambda_2} \, . \label{eq:3ptMHV}
\end{align}
It is easy to check that the above correlator satisfies BG equations for the three-gluon MHV amplitude.
The $z$-dependence is determined, as usual, by conformal invariance. As mentioned at the beginning of this section,  BG equations connect $z$-dependence with $\lambda$-dependence. It is highly nontrivial that they yield the correct $\lambda$-dependent prefactor.

\section{Background dilaton field}
The complex scalar dilaton field is defined as
\be\Phi=\frac{1}{\sqrt 2}(s+ia)\label{dild}\ee
where $s$ is the scalar dilaton field and $a$ is a pseudoscalar axion. We assume that it is a massless field.  We will be considering four-dimensional Dilaton-Yang-Mills (DYM) theory defined by the Lagrangian density
\be{\cal L}=\partial_\mu\Phi\partial^\mu\Phi^*-\frac{1}{2}\makebox{tr}F_{\mu\nu}F^{\mu\nu}-
\frac{1}{2}\Phi\,\makebox{tr}F_{-\mu\nu}F^{\mu\nu}_- -\frac{1}{2}\Phi^*\makebox{tr}F_{+\mu\nu}F^{\mu\nu}_+
+J^*\Phi+J\Phi^*\ ,\label{lagr}
\ee
where $F_+^{\mu\nu}$ and $F_-^{\mu\nu}$ are the purely selfdual and anti-selfdual gauge field strengths, respectively:
\be F_\pm^{\mu\nu}=\frac{1}{2}( F^{\mu\nu}\pm {\textstyle\frac{i}{2}}\epsilon^{\mu\nu\rho\sigma} F_{\rho\sigma})\ .
\label{selfd}\ee
We also introduced a classical source of the dilaton field,
\be J=\frac{1}{\sqrt 2}(J_s+iJ_a)\ .\label{source}\ee
This type of a scalar particle naturally accompanies the graviton in closed superstring theory. Since a shift of the real scalar $s$ by a constant value can be absorbed into the gauge coupling constant, the scattering amplitudes involving a number of gluons and one zero momentum dilaton reduce to pure Yang-Mills amplitudes modulo multiplicative constants.\footnote{The soft dilaton behavior was first discussed in
Refs.\cite{Ademollo:1975pf} and \cite{Shapiro:1975cz}.} However, the axion $a$ decouples at zero momentum. For non-zero momentum, Yang-Mills amplitudes with a single dilaton insertion can be obtained from Yang-Mills amplitudes in which the dilaton is replaced by two gluons with opposite helicities \cite{Stieberger:2016lng}. Here, however, we are interested in computing Yang-Mills amplitudes in the dilaton background field, {\em i.e}.\ in the case of off-shell dilatons.

We are interested in CCFT with a dilaton source localized on the celestial sphere. The corresponding two-dimensional action term is
\be S_\Phi=\int d^2z \big[\mu^*\Phi_{\Delta=2}^{+\epsilon}(z,\bz)+\mu^*\Phi_{\Delta=2}^{-\epsilon}(z,\bz)+
\mu\Phi_{\Delta=2}^{*+\epsilon }(z,\bz)+\mu\Phi_{\Delta=2}^{*-\epsilon }(z,\bz)\big],\label{scc1}\ee
where $\Phi_{\Delta=2}^{\pm\epsilon}(z,\bz)$ are the (complex) primary fields associated with incoming and outgoing dilatons, respectively, with dimensions $\Delta=2$ and zero spins. Here, $\mu=(\mu_s+i\mu_a)/\sqrt 2$ is a constant ``charge'' parameter.
We can write
\be\Phi_{\Delta=2}^{\pm\epsilon}(z,\bz)=\int d^4x \,\varphi^{\pm\epsilon}_{\Delta=2}(x;z)\Phi(x)\ ,\ee
where
\be\varphi^{\pm\epsilon}_{\Delta=2}(x;z)=\int_0^\infty d\omega\,\omega\, e^{\pm ip(\omega, z)\cdot x}\ , ~\quad p(\omega,z)=\omega(1+|z|^2,z+\bz,i(\bz-z),1-|z|^2)\ ,\ee
is the $\Delta=2$ bulk-to-boundary massless scalar propagator given by
the dimension $\Delta=2$ conformal primary wavefunction \cite{Pasterski1705}.
Now the two-dimensional action term (\ref{scc1})  reads as the four-dimensional source of the dilaton field,
\be S_\Phi=\int d^4x\big[J^*(x)\Phi(x)+J(x)\Phi^*(x)\big]\ ,\ee
with the current
\be J(x) ~=~ \mu\! \int d^2z\big[\varphi^{+\epsilon}_{\Delta=2}(x;z)+\varphi^{-\epsilon}_{\Delta=2}(x;z)\big] ~=~ \mu\!
\int d^4 Q\,\delta(Q^2)\, e^{iQ\cdot x}\ .
\ee
In the momentum space,
\be J(Q)=\mu(2\pi)^4\delta(Q^2)\label{conw2}\ .\ee

Our goal is to show that the single-valued amplitudes  are given by Mellin transforms of DYM gluon amplitudes, evaluated with the Lagrangian written in Eq.(\ref{lagr}),  with one insertion of the dilaton source (\ref{conw2}) of  the ``holomorphic'' $\Phi$ background.
These amplitudes are shown schematically in Fig.\ 1.
\begin{figure}[h]
\begin{center}
\begin{picture}(300,100)(0,-10)
\SetColor{Black}
\GCirc(150,30){22}{0.5}
\Gluon(133,45)(110,82){4}{3}
\LongArrow(135,65.0)(134.8,65.4)
\Gluon(143,51)(130,92){4}{3}
\LongArrow(121,58.0)(120.8,58.4)
\Gluon(167,45)(190,82){4}{3}
\LongArrow(158,67.0)(158.2,67.4)
\Gluon(157,51)(170,92){4}{3}
\LongArrow(171.8,60.0)(172.3,60.8)
\Gluon(140,10)(118,-23){4}{3}
\LongArrow(133,-5.0)(134.56,-3.33)
\Gluon(150,8)(150,-30){4}{3}
\LongArrow(153.6,-7.1)(153.6,-6.6)
\Gluon(160,10)(182,-23){4}{3}
\LongArrow(172,-2)(171.56,-1.33)
\SetColor{Red}
\DashArrowLine(260,30)(172,30){2.7}
\Text(210,20)[l]{$\color{red}Q$}
\BCirc(260,30){10}
\Line(254.5,38.5)(266.5,22)
\Line(254.5,22)(266,38.5)
\Text(280,40)[l]{$\color{red}J(Q)$}
\end{picture}
\end{center}
\caption{Gluon scattering in the background of dilaton field.}\label{4g1source}
\end{figure}
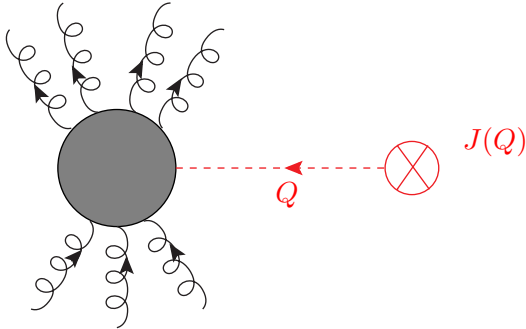

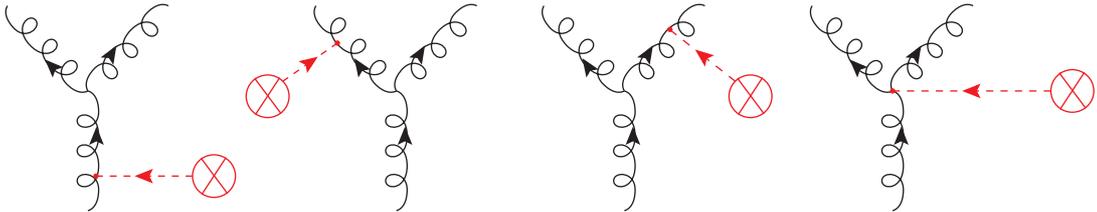
\begin{figure}[h]
\begin{center}
\begin{picture}(300,100)(0,-10)
\SetColor{Black}
\Gluon(-17,50)(-47,82){4}{3}
\LongArrow(-30.2,59)(-30.5,59.3)
\Gluon(-17,50)(13,82){4}{3}
\LongArrow(-8.5,63.5)(-8.2,64)
\Gluon(-17,50)(-17,5){4}{3}
\LongArrow(-13.8,32.5)(-13.8,33)

\Gluon(98,50)(68,82){4}{3}
\LongArrow(84.5,59)(84.3,59.3)
\Gluon(98,50)(128,82){4}{3}
\LongArrow(106.5,63.5)(106.8,64)
\Gluon(98,50)(98,5){4}{3}
\LongArrow(101.2,32.5)(101.2,33)

\Gluon(183,50)(153,82){4}{3}
\LongArrow(169.5,59)(169.3,59.3)
\Gluon(183,50)(213,82){4}{3}
\LongArrow(191.5,63.5)(191.8,64)
\Gluon(183,50)(183,5){4}{3}
\LongArrow(186.2,32.5)(186.2,33)

\Gluon(283,50)(253,82){4}{3}
\LongArrow(269.8,59)(269.5,59.3)
\Gluon(283,50)(313,82){4}{3}
\LongArrow(291.5,63.5)(291.8,64)
\Gluon(283,50)(283,5){4}{3}
\LongArrow(286.2,32.5)(286.2,33)

\SetColor{Red}
\DashArrowLine(22,18)(-14,18){2.7}
\Vertex(-14,18){1}
\BCirc(30,18){8}
\Line(25.2,24.7)(35,11.5)
\Line(25.5,11.2)(34.5,25.0)

\DashArrowLine(55.6,53.6)(76,68){2.7}
\Vertex(76,68){1}
\BCirc(50,48){8}
\Line(45.2,54.7)(55,41.5)
\Line(45.5,41.2)(54.5,55.0)

\DashArrowLine(224.4,53.6)(200,73){2.7}
\Vertex(200,73.1){1}
\BCirc(230,48){8}
\Line(225.2,54.7)(235,41.5)
\Line(225.5,41.2)(234.5,55.0)

\DashArrowLine(342,50)(283,50){2.7}
\Vertex(283,50){1}
\BCirc(350,50){8}
\Line(345.2,56.7)(355,43.5)
\Line(345.5,43.2)(354.5,57.0)
\end{picture}
\end{center}
\caption{Feynman diagrams for three gluons coupled to the dilaton source. }
\end{figure}
The source is connected to gluons through the scalar propagator, therefore the contribution from the respective part of Feynman diagrams (marked in red in Fig.\ 1) is given by
\be i(2\pi)^4{\delta(Q^2)\over Q^2}=-i(2\pi)^4\delta'(Q^2)\ ,\label{fcc}\ee
where we used $\delta(x)/x = -\delta'(x)$.
The remaining part is the scattering amplitude\footnote{More
precisely, it is the so-called invariant matrix element as obtained from Feynman diagrams, {\it i.e}.\ the amplitude without the momentum-conserving delta function.}
involving a number of gluons and one off-shell dilaton $\Phi$ with momentum $Q$. As an example, we show in Fig.\ 2 the diagrams contributing to three-gluon amplitudes. Similar processes have been studied in the past
in the context of effective theory describing massive Higgs bosons and axions coupled to gluons in the standard model \cite{Dixon:2004za}. The MHV (partial) amplitudes are given by
\be {\cal M}(1^-,2^-, 3^+,\cdots, N^+)_{\Phi(Q)}=\frac{\langle 12\rangle^4}{\langle 12\rangle\langle 23\rangle\cdots\langle N1\rangle}\ ,\label{mhvd}
\ee
with the PT denominator determined by the group factor \cite{Taylor:2017sph}. Although this looks the same as the Parke-Taylor formula \cite{Parke:1986gb}, the momentum conservation works in a different way because $\sum p_{out}-\sum p_{in}=Q$, therefore the dilaton supplies momentum to the gluon system. Furthermore,
\be {\cal M}(1^-,2^-, 3^+,\cdots, N^+)_{\Phi^*(Q)}=0\ ,\ee
and
\begin{align}
{\cal M}(1^+,2^+, 3^-,\cdots, N^-)_{\Phi(Q)} &=0\ ,\\
{\cal M}(1^+,2^+, 3^-,\cdots, N^-)_{\Phi^*(Q)} &=\frac{[ 12]^4}{[ 12][ 23]\cdots [N1]}\ ,\label{mhvbd}
\end{align}
therefore $\Phi$ couples to MHV amplitudes and projects out the $\overline{\rm MHV}$ sector, while $\Phi^*$ projects out MHV amplitudes and couples to the $\overline{\rm MHV}$ sector, as expected from their couplings to the anti-selfdual and selfdual gauge sectors, respectively. There are also some amplitudes that are non-zero for off-shell dilatons only, for example
\be {\cal M}(1^+,2^+, 3^+,\cdots, N^+)_{\Phi(Q)}=\frac{(Q^2)^2}{\langle 12\rangle\langle 23\rangle\cdots\langle N1\rangle}\ .\ee
Due to the factor (\ref{fcc}), these amplitudes vanish, however, in the background under consideration.
\subsection{Three-gluon correlators}
We want to show that the three-gluon correlator
(\ref{eq:3ptMHV}) is given by the Mellin transform of  the three-gluon amplitude evaluated in the dilaton background (\ref{conw2}):
\begin{align}
&\Big\langle \phi_{\D_1,-}^{a_1,-\epsilon}(z_{1},\zbar_{1})\, \phi_{\D_2,-}^{a_2,-\epsilon}(z_2,\bz_2)\,\phi_{\D_3,+}^{a_3,+\epsilon}(z_3,\bz_3)\Big\rangle_{\Phi} \nonumber\\
&=~f^{a_1a_2a_3}\int_0^{\infty} d\omega_1 \int_0^{\infty} d\omega_2 \int_0^{\infty} d\omega_3\, \omega_1^{i\lambda_1}\omega_2^{i\lambda_2}\omega_3^{i\lambda_3}
\mathcal{M}(1^-,2^-,3^+)_{\Phi(Q)}
\frac{\delta(Q^2)}{Q^2} \, , \label{eq:3ptMellin}
\end{align}
where, according to Eq.(\ref{mhvd}),
\be
\mathcal{M}(1^-,2^-,3^+)_{\Phi(Q)} =  \frac{\langle 12\rangle^3}{\langle 23 \rangle \langle 31 \rangle} = \frac{\omega_1\omega_2}{\omega_3} \frac{z_{12}^3}{z_{23}z_{31}}\, .
\ee

Since
\be
Q^2 = (p_1+p_2-p_3)^2 = \omega_1\omega_2 z_{12}\bz_{12}- \omega_1\omega_3 z_{13}\bz_{13}-\omega_2\omega_3 z_{23}\bz_{23} \, ,
\ee
the current-propagator factor is given by
\begin{align}
\frac{\delta(Q^2)}{Q^2}& = \frac{\delta(\omega_1\omega_2 z_{12}\bz_{12}- \omega_1\omega_3 z_{13}\bz_{13}-\omega_2\omega_3 z_{23}\bz_{23}) }{\omega_1\omega_2 z_{12}\bz_{12}- \omega_1\omega_3 z_{13}\bz_{13}-\omega_2\omega_3 z_{23}\bz_{23} } \nonumber\\
&=\,-\frac{\delta\Big(\omega_3-\frac{\omega_1\omega_2 z_{12}\bz_{12}}{\omega_1 z_{13} \bz_{13}+\omega_2 z_{23}\bz_{23}}\Big)}{(\omega_1 z_{13}\bz_{13}+\omega_2 z_{23}\bz_{23})^2\Big( \omega_3-\frac{\omega_1\omega_2 z_{12}\bz_{12}}{\omega_1 z_{13} \bz_{13}+\omega_2 z_{23}\bz_{23}}\Big)} \nonumber\\
&= ~~\frac{\delta'\Big(\omega_3-\frac{\omega_1\omega_2 z_{12}\bz_{12}}{\omega_1 z_{13} \bz_{13}+\omega_2 z_{23}\bz_{23}}\Big)}{(\omega_1 z_{13}\bz_{13}+\omega_2 z_{23}\bz_{23})^2} \, ,
\end{align}
where we used $\delta(x)/x = -\delta'(x)$. After integrating this delta function derivative over $\omega_3$, the r.h.s.\  of Eq.(\ref{eq:3ptMellin}) becomes
\begin{align}
&(i\lambda_3-1)\frac{z_{12}^3}{z_{23}z_{13}} (z_{12}\bz_{12})^{i\lambda_3-2}\!\int_0^{\infty}\!\! d\omega_1\!\! \int_0^{\infty} \!\!d\omega_2\, \omega_1^{i\lambda_1+i\lambda_3-1}\, \omega_2^{i\lambda_2+i\lambda_3-1} (\omega_1 z_{13}\bz_{13} +\omega_2 z_{23} \bz_{23})^{-i\lambda_3}\nonumber\\
&\qquad =~(i\lambda_3-1)\frac{z_{12}^3}{z_{23}z_{13}} (z_{12}\bz_{12})^{i\lambda_3-2} (z_{23}\bz_{23})^{-i\lambda_2-i\lambda_3} (z_{13}\bz_{13})^{i\lambda_2}\, B(-i\lambda_2,i\lambda_2+i\lambda_3)\\ &~~~~~~~~~~~~~~~\qquad\qquad\qquad\qquad\times \int_0^{\infty} d\omega_1 \omega_1^{i\lambda_1+i\lambda_2+i\lambda_3-1}\nonumber\end{align}
After performing the last integration, we obtain
\begin{align}
f^{a_1a_2a_3}&\int_0^{\infty} d\omega_1 \int_0^{\infty} d\omega_2 \int_0^{\infty} d\omega_3\, \omega_1^{i\lambda_1}\omega_2^{i\lambda_2}\omega_3^{i\lambda_3}
\mathcal{M}(1^-,2^-,3^+)_{\Phi(Q)}
\frac{\delta(Q^2)}{Q^2} \nonumber\\[1mm]
& =- 2\pi f^{a_1a_2a_3}  \delta(\lambda_1+\lambda_2+\lambda_3)(1+i\lambda_1+i\lambda_2)B(-i\lambda_1,-i\lambda_2)\nonumber\\[1mm] & ~~~~~~~~~~~~~~~\times z_{12}^{\, 1-i\lambda_1-i\lambda_2} z_{23}^{\, i\lambda_1-1}z_{13}^{\, i\lambda_2-1} \bz_{12}^{\, -i\lambda_1-i\lambda_2-2} \bz_{23}^{\, i\lambda_1} \bz_{13}^{\, i\lambda_2} \, ,
\end{align}
which matches Eq.(\ref{eq:3ptMHV}) up to an overall normalization factor,\footnote{In this context, ``normalization factors''  are numerical constants that depend neither on $z$'s nor on $\lambda$'s.}
therefore
\begin{align} \Big\langle \phi_{\D_1,-}^{a_1,-\epsilon}&(z_{1},\zbar_{1})\, \phi_{\D_2,-}^{a_2,-\epsilon}(z_2,\bz_2)\,\phi_{\D_3,+}^{a_3,+\epsilon}(z_3,\bz_3)\Big\rangle_{\rm SV}\nonumber\\ &=
\Big\langle \phi_{\D_1,-}^{a_1,-\epsilon}(z_{1},\zbar_{1})\, \phi_{\D_2,-}^{a_2,-\epsilon}(z_2,\bz_2)\,\phi_{\D_3,+}^{a_3,+\epsilon}(z_3,\bz_3)\Big\rangle_{\Phi}\ . \end{align}

\subsection{Four-gluon correlators}
We want to show that the four-point correlator is given by
\begin{align}
\Big\langle\phi_{\D_1,-}^{a_1,-\epsilon}&(z_{1},\zbar_{1}
)\,
\phi_{\D_2,-}^{a_2,-\epsilon}(z_2,\bz_2)
\,\phi_{\D_3,+}^{a_3,+\epsilon}
(z_3,\bar z_3)
\,\phi_{\D_4,+}^{a_4,+\epsilon}(z_4,\bz_4)\Big\rangle_{\Phi}= \nonumber\\[1mm] &~~=f^{a_1a_2b}f^{a_3a_4b}M(z_i,\bz_i)_{\Phi}
+f^{a_1a_3b}f^{a_2a_4b} \widetilde M(z_i,\bz_i)_{\Phi}\ ,\label{partials1}
\end{align}
with \begin{align}
M(z_i,\bz_i)_{\Phi}=\int_0^{\infty}d\omega_1 &\!\int_0^{\infty} d\omega_2 \!\int_0^{\infty} d\omega_3 \!\int_0^{\infty} d\omega_4 \, \omega_1^{i\lambda_1}\omega_2^{i\lambda_2}\omega_3^{i\lambda_3}\omega_4^{i\lambda_4}\nonumber\\ &\times
\mathcal{M}(1^-,2^-,3^+,4^+)_{\Phi(Q)}\frac{\delta(Q^2)}{Q^2} \, , \label{eq:4ptMellin}
\end{align}
and a similar expression for $\widetilde M(z_i,\bz_i)_{\Phi}$, with $\mathcal{M}(1^-,3^+,2^-,4^+)_{\Phi(Q)}$.

According to Eq.(\ref{mhvd}),
\be
\mathcal{M}(1^-,2^-,3^+,4^+)_{\Phi(Q)} =  \frac{\langle 12\rangle^3}{\langle 23 \rangle \langle 34 \rangle\langle 41\rangle} = \frac{\omega_1\omega_2}{\omega_3\omega_4} \frac{z_{12}^3}{z_{23}z_{34}z_{41}}\, .
\ee
In this case,
\begin{align}
Q^2 &= (p_1+p_2-p_3-p_4)^2  \nonumber\\
&= \omega_1\omega_2 z_{12}\bz_{12}- \omega_1\omega_3 z_{13}\bz_{13}-\omega_2\omega_3 z_{23}\bz_{23}-\omega_1\omega_4 z_{14}\bz_{14}-\omega_2\omega_4 z_{24}\bz_{24}+\omega_3\omega_4 z_{34}\bz_{34} \, .
\end{align}
Now the current-propagator factor reads
\begin{align}
\frac{\delta(Q^2)}{Q^2} =& \frac{\delta(\omega_1-\omega_1^*) }{(\omega_2 z_{12}\bz_{12} -\omega_3 z_{13}\bz_{13}-\omega_4 z_{14}\bz_{14})^2(\omega_1-\omega_1^*)}  \nonumber\\
=& -\frac{\delta'(\omega_1-\omega_1^*)}{(\omega_2 z_{12}\bz_{12} -\omega_3 z_{13}\bz_{13}-\omega_4 z_{14}\bz_{14})^2} \, ,
\end{align}
where
\be
\omega_1^* = \frac{\omega_2\omega_3 z_{23}\bz_{23} +\omega_2\omega_4 z_{24}\bz_{24}-\omega_3\omega_4 z_{34}\bz_{34}}{ \omega_2 z_{12}\bz_{12} -\omega_3 z_{13}\bz_{13}-\omega_4 z_{14}\bz_{14}} \, .
\ee
After integrating the above delta function derivative over $\omega_1$,
Eq.(\ref{eq:4ptMellin}) becomes
\begin{align}
& M(z_i,\bz_i)_{\Phi}=  \int d\omega_2\int_0^{\infty} d\omega_3 \int_0^{\infty} d\omega_4 \, (1+i\lambda_1) \omega_2^{i\lambda_2+1}\omega_3^{i\lambda_3-1}\omega_4^{i\lambda_4-1} \frac{z_{12}^3}{z_{23}z_{34}z_{41}} \nonumber\\
&\times \left(\frac{\omega_2\omega_3 z_{23}\bz_{23} +\omega_2\omega_4 z_{24}\bz_{24}-\omega_3\omega_4 z_{34}\bz_{34}}{ \omega_2 z_{12}\bz_{12} -\omega_3 z_{13}\bz_{13}-\omega_4 z_{14}\bz_{14}}\right)^{i\lambda_1} (\omega_2 z_{12}\bz_{12} -\omega_3 z_{13}\bz_{13}-\omega_4 z_{14}\bz_{14})^{-2} \, . \label{eq:4ptMellinStep1}
\end{align}
We have not specified yet the integration region for $\omega_2$ because it is constrained by the positivity of $\omega_1^*$:
\be
\omega_1^* =\frac{\omega_2\omega_3 z_{23}\bz_{23} +\omega_2\omega_4 z_{24}\bz_{24}-\omega_3\omega_4 z_{34}\bz_{34}}{ \omega_2 z_{12}\bz_{12} -\omega_3 z_{13}\bz_{13}-\omega_4 z_{14}\bz_{14}} >0 \, .
\ee
This leads to
\begin{align}
&\omega_2> \frac{\omega_3 z_{13}\bz_{13} +\omega_4 z_{14}\bz_{14}}{z_{12}\bz_{12}} = a \\
\text{or } \,~ &0<\omega_2<\frac{\omega_3\omega_4 z_{34}\bz_{34}}{\omega_3 z_{23}\bz_{23} +\omega_4 z_{24} \bz_{24}} = b\ .
\end{align}
Then Eq.(\ref{eq:4ptMellinStep1}) becomes
\begin{align}
&M(z_i,\bz_i)_{\Phi}= \left(\int_0^b d\omega_2 +\int_a^{+\infty} d\omega_2\right)\int_0^{\infty} d\omega_3 \int_0^{\infty} d\omega_4 \, (1+i\lambda_1) \omega_2^{i\lambda_2+1}\omega_3^{i\lambda_3-1}\omega_4^{i\lambda_4-1} \frac{z_{12}^3}{z_{23}z_{34}z_{41}} \nonumber\\
&\times \left(\frac{\omega_2\omega_3 z_{23}\bz_{23} +\omega_2\omega_4 z_{24}\bz_{24}-\omega_3\omega_4 z_{34}\bz_{34}}{ \omega_2 z_{12}\bz_{12} -\omega_3 z_{13}\bz_{13}-\omega_4 z_{14}\bz_{14}}\right)^{i\lambda_1} (\omega_2 z_{12}\bz_{12} -\omega_3 z_{13}\bz_{13}-\omega_4 z_{14}\bz_{14})^{-2}  \nonumber\\[1mm]
&=  \frac{z_{12}^3}{z_{23}z_{34}z_{41}} \Bigg [ \int_0^{\infty} d\omega_3\int_0^{\infty} d\omega_4  B(-i\lambda_1,-i\lambda_2)(1+i\lambda_1+i\lambda_2) \, {\cal I}_1(\omega_3,\omega_4,z_i,\bz_i) \nonumber\\
&\qquad + B(2+i\lambda_1,2+i\lambda_2)(3+i\lambda_1+i\lambda_2)\, {\cal I}_2(\omega_3,\omega_4,z_i,\bz_i) \Bigg] \, , \label{eq:4ptMellinStep2}
\end{align}
where
\begin{align}
&{\cal I}_1(\omega_3,\omega_4,z_i,\bz_i)
=\omega_3^{i\lambda_3-1}\omega_4^{i\lambda_4-1} (z_{12} \bz_{12})^{-2-i\lambda_1-i\lambda_2} (\omega_3 z_{23} \bz_{23} +\omega_4 z_{24}\bz_{24})^{i\lambda_1} \nonumber\\
&\times (\omega_3 z_{13}\bz_{13}+\omega_4 z_{14}\bz_{14})^{i\lambda_2}\, _2F_1\left({-i\lambda_1,-i\lambda_2 \atop -1-i\lambda_1-i\lambda_2}; \frac{\omega_3\omega_4 z_{34}\bz_{34} z_{12}\bz_{12}}{(\omega_3 z_{23} \bz_{23} +\omega_4 z_{24}\bz_{24})(\omega_3 z_{13}\bz_{13}+\omega_4 z_{14}\bz_{14})} \right) \, ,
\end{align}
\begin{align}
&{\cal I}_2(\omega_3,\omega_4,z_i,\bz_i)
=\omega_3^{i\lambda_3-1}\omega_4^{i\lambda_4-1} (\omega_3\omega_4 z_{34}\bz_{34})^{2+i\lambda_1+i\lambda_2} (\omega_3 z_{23} \bz_{23} +\omega_4 z_{24}\bz_{24})^{-2-i\lambda_2} \nonumber\\
&\times(\omega_3 z_{13}\bz_{13}+\omega_4 z_{14}\bz_{14})^{-2-i\lambda_1}\, _2F_1\left( {2+i\lambda_1, 2+i\lambda_2 \atop 3+i\lambda_1+i\lambda_2};\frac{\omega_3\omega_4 z_{34}\bz_{34} z_{12}\bz_{12}}{(\omega_3 z_{23} \bz_{23} +\omega_4 z_{24}\bz_{24})(\omega_3 z_{13}\bz_{13}+\omega_4 z_{14}\bz_{14})} \right).
\end{align}
The best way of making contact with the results of Section 2 is by computing the function
\begin{align}
&G(x,\bar{x})_\Phi =\lim_{z_1,\bz_1\rightarrow \infty} z_1^{2h_1}\bz_1^{2\bar{h}_1} M[z_1;z_2=1,z_3=x,z_4=0]_\Phi \nonumber\\
=&-\frac{1}{x(1-x)} \Bigg[ \int_0^\infty d\omega_3 \int_0^{\infty} d\omega_4 B(-i\lambda_1,-i\lambda_2)(1+i\lambda_1+i\lambda_2) \omega_3^{i\lambda_3-1}\omega_4^{i\lambda_4-1} (\omega_3(1-x)(1-\bar{x})+\omega_4)^{i\lambda_1} \nonumber\\
&\quad\quad\quad \qquad ~~\times (\omega_3+ \omega_4)^{i\lambda_2} \, _2F_1\left({-i\lambda_1,-i\lambda_2 \atop -1-i\lambda_1-i\lambda_2}; \frac{\omega_3\omega_4 \, x\, \bar{x}}{(\omega_3 (1-x)(1-\bar{x}) +\omega_4 )(\omega_3 +\omega_4 )} \right) \nonumber\\
& + \int_0^\infty d\omega_3 \int_0^{\infty} d\omega_4 B(2+i\lambda_1,2+i\lambda_2)(3+i\lambda_1+i\lambda_2)\omega_3^{i\lambda_3-1}\omega_4^{i\lambda_4-1} (\omega_3\omega_4 \, x \, \bar{x})^{2+i\lambda_1+i\lambda_2} (\omega_3+\omega_4)^{-2-i\lambda_1}\nonumber\\
& \times (\omega_3(1-x)(1-\bar{x})+\omega_4)^{-2-i\lambda_2} \, _2F_1\left( {2+i\lambda_1, 2+i\lambda_2 \atop 3+i\lambda_1+i\lambda_2}; \frac{\omega_3\omega_4 \, x\, \bar{x}}{(\omega_3 (1-x)(1-\bar{x}) +\omega_4 )(\omega_3 +\omega_4 )}\right) \Bigg] \, . \label{eq:G1_Int1}
\end{align}
At this point, we change the integration variables to $\omega_p = \omega_3+\omega_4, ~ t = {\omega_3}/{\omega_p}$ and perform the integral over $\omega_p$, with the result:\footnote{Here again, we skipped an irrelevant normalization factor.}
\begin{align}
&G(x,\bar{x})_\Phi =\delta(\lambda_1+\lambda_2+\lambda_3+\lambda_4)\nonumber\\
&\times \frac{1}{x(1-x)} \Bigg[ \int_0^1 dt\, B(-i\lambda_1,-i\lambda_2)(1+i\lambda_1+i\lambda_2)\, t^{i\lambda_3-1}(1-t)^{i\lambda_4-1} \big(t(1-x)(1-\bar{x})+1-t\big)^{i\lambda_1} \nonumber\\
&\qquad \qquad \qquad\times \, _2F_1\left({-i\lambda_1,-i\lambda_2 \atop -1-i\lambda_1-i\lambda_2}; \frac{t(1-t) \, x\, \bar{x}}{t (1-x)(1-\bar{x}) +1-t } \right) \nonumber\\
&\qquad \qquad \qquad+\int_0^1 dt \, B(2+i\lambda_1,2+i\lambda_2)(3+i\lambda_1+i\lambda_2)\, t^{1-i\lambda_4}(1-t)^{1-i\lambda_3} (x\bar{x})^{2+i\lambda_1+i\lambda_2}\nonumber\\
&\qquad \qquad \qquad\times\big(t(1-x)(1-\bar{x})+1-t\big)^{-2-i\lambda_2}\, _2F_1\left( {2+i\lambda_1, 2+i\lambda_2 \atop 3+i\lambda_1+i\lambda_2}; \frac{t(1-t) \, x\, \bar{x}}{t (1-x)(1-\bar{x}) +1-t }\right) \Bigg] \, . \label{eq:G1_IntRep}
\end{align}
This looks quite different from the series expansions obtained in Section 2. For small $|x|$, however, we can
expand it in a power series of $x$ and $\bar{x}$ and perform the remaining integrals over $t$ analytically.
We checked that this series expansion matches the
$s$-channel expansion of Eq.(\ref{eq:sol_G1_sv}), which is sufficient to prove that the four-gluon celestial amplitude evaluated in the dilaton background is equal to the correlator obtained by solving the BG equations.\footnote{The integral (\ref{eq:G1_IntRep}) and series (\ref{eq:sol_G1_sv}) representations of the four-gluon single-valued amplitude will be studied in more detail in Ref.\cite{prep}.}
Note that the relation
\be \widetilde G(x,\bar{x})_\Phi =-x\, G(x,\bar{x})_\Phi\ee
follows immediately because
\be \mathcal{M}(1^-,3^+,2^-,4^+)_{\Phi(Q)}=-x
\mathcal{M}(1^-,2^-,3^+,4^+)_{\Phi(Q)}\ .\ee
In this way, we conclude that
\begin{align}
\Big\langle\phi_{\D_1,-}^{a_1,-\epsilon}&(z_{1},\zbar_{1}
)\,
\phi_{\D_2,-}^{a_2,-\epsilon}(z_2,\bz_2)
\,\phi_{\D_3,+}^{a_3,+\epsilon}
(z_3,\bar z_3)
\,\phi_{\D_4,+}^{a_4,+\epsilon}(z_4,\bz_4)\Big\rangle_{\rm SV} \nonumber\\[1mm] &=
\Big\langle\phi_{\D_1,-}^{a_1,-\epsilon}(z_{1},\zbar_{1}
)\,
\phi_{\D_2,-}^{a_2,-\epsilon}(z_2,\bz_2)
\,\phi_{\D_3,+}^{a_3,+\epsilon}
(z_3,\bar z_3)
\,\phi_{\D_4,+}^{a_4,+\epsilon}(z_4,\bz_4)\Big\rangle_{\Phi}
\ .
\end{align}

\section{Leading OPEs and the MHV projection}
\subsection{Leading OPEs}
In this section, we will extract the leading OPE terms from the four-gluon single-valued amplitude written in Eq.(\ref{eq:full_MHV_SV}). These are the products of gluon operators fusing into primary fields with dimensions having the lowest integer part, that is with $\Delta=1+ i\lambda$.

In the past, the leading OPE terms have been extracted in two ways. The one closest to the procedure followed in this paper, is by studying the collinear limits of Mellin-transformed amplitudes
\cite{Fan1903}. The second one utilizes recursion relations implied by the asymptotic symmetries of CCFT, in particular by the (super)translational invariance \cite {Strominger1910}. A priori, there is no reason to expect that the single-valued amplitude (\ref{eq:full_MHV_SV}) yields the same OPEs because  it is evaluated in the dilaton background, and supertranslational invariance is broken by the momentum supplied to the gluon system by the dilaton field. However, the BG equations, satisfied by the amplitude (\ref{eq:full_MHV_SV}), are based on the  OPEs of Refs.\cite{Fan1903,Strominger1910}. Furthermore, the dilaton couples to anti-selfdual fields only, therefore the selfdual part, in particular the products involving positive helicities only, should remain unaffected. For these reasons, we expect that the new OPEs will be related to the  OPEs of Refs.\cite{Fan1903,Strominger1910} -- and we will see how they are related indeed.

We start from the $s$-channel expansions of Section 2.1. The limit $x\to 0$ allows studying $z_1\to z_2$ and $z_3\to z_4$.
The dominant contribution to the single-valued amplitude is written
in Eq.(\ref{eq:z34limit}). From $z_3\to z_4$, we obtain the familiar \cite{Fan1903,Strominger1910}
\begin{align}
\phi_{\D_3,+}^{a_3,+\epsilon}(z_3,\bar z_3)\,\phi_{\D_4,+}^{a_4,+\epsilon}(z_4,\bz_4) = \frac{-i f^{a_3 a_4 x}}{z_{34}} B(\Delta_3-1,\Delta_4-1)\phi_{\D_3+\D_4-1,+}^{x,+\epsilon}(z_3,\bz_3) \, . \label{eq:z34OPE}
\end{align}
In the limit $z_1\to z_2$, however, we obtain
\begin{align}
\phi_{\D_1,-}^{a_1,-\epsilon}(z_1,\bar z_1)\,\phi_{\D_2,-}^{a_2,-\epsilon}(z_2,\bz_2) &= i f^{a_1 a_2 x} \,\frac{z_{12}}{\bz_{12}^2}\, (z_{12}\bz_{12})^{2-\Delta_1-\Delta_2}\nonumber\\ &~~~\times \! (1-\Delta_1-\Delta_2)B(1-\Delta_1,1-\Delta_2)\widetilde\phi_{3-\D_1-\D_2,+}^{x,-\epsilon}(z_2,\bz_2) \, . \label{eq:z12OPE}\end{align}
This is different from the usual OPE which yields a {\it negative\/}
helicity field $\phi_{\D_1+\D_2-1,-}^{x,-\epsilon}$ and  a simple $\bz_{12}^{-1}$ pole \cite{Fan1903,Strominger1910}. The field denoted by $\widetilde\phi$ has dimension $3-\D_1-\D_2=1-i\lambda_1-i\lambda_2$ and spin $J=+1$, which are exactly the dimension and spin of the shadow of  $\phi_{\D_1+\D_2-1,-}^{x,-\epsilon}$. Similar operators will appear in other OPEs. It is unlikely though that nonlocal, shadow-transformed operators appear in the OPEs of local operators. For this reason, we will call them ``quasishadows.''
In the second part of this section, we will gain some insight into the properties of quasishadow fields.

The limit $x\to 1$ allows studying $z_2\to z_3$ and $z_1\to z_4$. In this case, both partial amplitudes contribute to leading OPE terms and can be combined by using Jacobi identities for group factors. The $u$-channel expansions of Section 2.2 yield
\begin{align}
\phi_{\D_2,-}^{a_2,-\epsilon}(z_2,\bz_2)\,\phi_{\D_3,+}^{a_3,+\epsilon}(z_3,\bar z_3) = \frac{-i f^{a_2a_3 x}}{z_{23}}\Big[& B(\D_3-1,1-\D_2-\D_3)\phi_{\D_2+\D_3-1,-}^{x,-\epsilon}(z_3,\bz_3) \nonumber\\
&-B(\D_2+1,1-\D_2-\D_3)\phi_{\D_2+\D_3-1,-}^{x,+\epsilon}(z_3,\bz_3) \Big] \, ,\nonumber\\
+~\frac{i f^{a_2a_3x}}{z_{23}}(z_{23}\bar{z}_{23})^{2-\Delta_2-\Delta_3}\Big[ (3-\Delta_2-&\Delta_3)B(1-\Delta_2,\Delta_2+\Delta_3-3)
\widetilde\phi_{3-\Delta_2-\Delta_3,-}^{x,+\epsilon}(z_3,\bar{z}_3)\nonumber\\
-(3-\Delta_2&-\Delta_3)B(3-\Delta_3,\Delta_2+\Delta_3-3)
\widetilde\phi_{3-\Delta_2-\Delta_3,-}^{x,-\epsilon}(z_3,\bar{z}_3) \Big] \, , \label{eq:z23OPE} \\
\phi_{\D_1,-}^{a_1,-\epsilon}(z_1,\bz_1)\,\phi_{\D_4,+}^{a_4,+\epsilon}(z_4,\bar z_4) = \frac{-i f^{a_1a_4 x}}{z_{14}}\Big[& B(\D_4-1,1-\D_1-\D_4)\phi_{\D_1+\D_4-1,-}^{x,-\epsilon}(z_4,\bz_4) \nonumber\\
&-B(\D_1+1,1-\D_1-\D_4)\phi_{\D_1+\D_4-1,-}^{x,+\epsilon}(z_4,\bz_4) \Big]\nonumber\\
+~\frac{i f^{a_1a_4x}}{z_{14}}{(z_{14}\bar{z}_{14})^{2-\Delta_1-\Delta_4}}\Big[ (3-\Delta_1-&\Delta_4)B(1-\Delta_1,\Delta_1+\Delta_4-3)\widetilde\phi_{3-\Delta_1-\Delta_4,-}^{x,+\epsilon}
(z_4,\bar{z}_4)\nonumber\\
-(3-\Delta_1&-\Delta_4)B(3-\Delta_4,\Delta_1+\Delta_4-3)
\widetilde\phi_{3-\Delta_1-\Delta_4,-}^{x,-\epsilon}(z_4,\bar{z}_4) \Big]  \, . \label{eq:z14OPE}
\end{align}
As expected, both $z_2\to z_3$ and $z_1\to z_4$ limits yield the same OPE.
Similarly, the $t$-channel expansions of Section 2.3, valid at $1/x\approx 0$,  yield
\begin{align}
\phi_{\D_1,-}^{a_1,-\epsilon}(z_1,\bz_1)\,\phi_{\D_3,+}^{a_3,+\epsilon}(z_3,\bar z_3) = \frac{-i f^{a_1a_3 x}}{z_{13}}\Big[& B(\D_3-1,1-\D_1-\D_3)\phi_{\D_1+\D_3-1,-}^{x,-\epsilon}(z_3,\bz_3) \nonumber\\
&-B(\D_1+1,1-\D_1-\D_3)\phi_{\D_1+\D_3-1,-}^{x,+\epsilon}(z_3,\bz_3) \Big] \nonumber\\
+~\frac{i f^{a_1a_3x}}{z_{13}}{(z_{13}\bar{z}_{13})^{2-\Delta_1-\Delta_3}}\Big[ (3-\Delta_1-&\Delta_3)B(1-\Delta_1,\Delta_1+\Delta_3-3)\widetilde\phi_{3-\Delta_1-\Delta_3,-}^{x,+\epsilon}
(z_3,\bar{z}_3)\nonumber\\
-(3-\Delta_1&-\Delta_3)B(3-\Delta_3,\Delta_1+\Delta_3-3)\widetilde\phi_{3-\Delta_1-\Delta_3,-}^{x,-\epsilon}
(z_3,\bar{z}_3) \Big]  \, , \label{eq:z13OPE} \\
\phi_{\D_2,-}^{a_2,-\epsilon}(z_2,\bz_2)\,\phi_{\D_4,+}^{a_4,+\epsilon}(z_4,\bar z_4) = \frac{-i f^{a_2a_4 x}}{z_{24}}\Big[& B(\D_4-1,1-\D_2-\D_4)\phi_{\D_2+\D_4-1,-}^{x,-\epsilon}(z_4,\bz_4) \nonumber\\
&-B(\D_2+1,1-\D_2-\D_4)\phi_{\D_2+\D_4-1,-}^{x,+\epsilon}(z_4,\bz_4) \Big] \nonumber\\
+~\frac{i f^{a_2a_4x}}{z_{24}}{(z_{24}\bar{z}_{24})^{2-\Delta_2-\Delta_4}}\Big[ (3-\Delta_2-&\Delta_4)B(1-\Delta_2,\Delta_2+\Delta_4-3)\widetilde\phi_{3-\Delta_2-\Delta_4,-}^{x,+\epsilon}
(z_4,\bar{z}_4)\nonumber\\
-(3-\Delta_2&-\Delta_4)B(3-\Delta_4,\Delta_2+\Delta_4-3)\widetilde\phi_{3-\Delta_2-\Delta_4,-}^{x,-\epsilon}
(z_4,\bar{z}_4) \Big] \, . \label{eq:z24OPE}
\end{align}
We see that the OPEs derived from $t$-channel expansions agree with the $u$-channel OPEs, in agreement with the symmetry of the amplitude under $3\leftrightarrow 4$, which is a necessary condition for the crossing symmetry of the conformal correlator. All of them are equivalent to Eq.(\ref{eq:z23OPE}).

By looking at Eqs.(\ref{eq:z34OPE}), (\ref{eq:z12OPE}) and (\ref{eq:z23OPE}), we observe that the OPE terms singular as $z_{ij}^{-1}$ agree with the OPEs written in Refs.\cite{Fan1903,Strominger1910}. However, the terms singular as  $\bz_{ij}^{-1}$  are replaced here by the terms involving quasishadow fields, singular as $|z_{ij}|^{-1}$, but with nontrivial phase factors. We will describe the deformation of OPEs occurring in the presence of the dilaton background as the effect of ``MHV projection.'' But first, we need to discuss some properties of the quasishadow fields.
\subsection{Quasishadow correlators}
The quasishadow field first appeared in the OPE (\ref{eq:z12OPE}), extracted from the $x\to 0$ limit of the single-valued amplitude written in Eq.(\ref{eq:z34limit}). From the same equation, we can also extract the three-point function of this field with two gluon operators:
\begin{align}
\Big\langle\widetilde{\phi}_{\widetilde{\D}_2,+}^{a_2,-\epsilon}&(z_2,\bz_2)
\,\phi_{\D_3,+}^{a_3,+\epsilon}
(z_3,\bar z_3)
\,\phi_{\D_4,+}^{a_4,+\epsilon}(z_4,\bz_4)\Big\rangle_{\text{SV}}  \nonumber\\ \label{qss1}
&=i f^{a_2a_3a_4}  B(i\lambda_3,i\lambda_4)\delta(\lambda_2+\lambda_3+\lambda_4)\, z_{34}^{-1} z_{23}^{-i\lambda_3-1} \bz_{23}^{-i\lambda_3}z_{24}^{-i\lambda_4-1}\bz_{24}^{-i\lambda_4} \, ,
\end{align}
where $\widetilde{\D}_2=1-i\lambda_2=1+i\lambda_3+i\lambda_4$. We want to show that this amplitude is identical to the shadow-transformed $\overline{\text{MHV}}$  amplitude
\begin{align}
&\Big\langle\widetilde{\phi_{\D_2,-}^{a_2,-\epsilon}}(z_2',\bz_2')
\,\phi_{\D_3,+}^{a_3,+\epsilon}
(z_3,\bar z_3)
\,\phi_{\D_4,+}^{a_4,+\epsilon}(z_4,\bz_4)\Big\rangle_{\Phi^*} \nonumber\\
&=\int \frac{d^2 z_2}{(z_2-z_2')^{2-i\lambda_2}(\bz_2-\bz_2')^{-i\lambda_2}}
\Big\langle\phi_{\D_2,-}^{a_2,-\epsilon}(z_2,\bz_2)
\,\phi_{\D_3,+}^{a_3,+\epsilon}
(z_3,\bar z_3)
\,\phi_{\D_4,+}^{a_4,+\epsilon}(z_4,\bz_4)\Big\rangle_{\Phi^*}
\label{shad1}
\end{align}
with
\begin{align}
&\Big\langle\phi_{\D_2,-}^{a_2,-\epsilon}(z_2,\bz_2)
\,\phi_{\D_3,+}^{a_3,+\epsilon}
(z_3,\bar z_3)
\,\phi_{\D_4,+}^{a_4,+\epsilon}(z_4,\bz_4)\Big\rangle_{\Phi^*}
 \nonumber\\
&=f^{a_2a_3a_4}\int_0^{\infty} d\omega_2 \int_0^{\infty} d\omega_3\int_0^{\infty} d\omega_4\, \omega_2^{i\lambda_2}\omega_3^{i\lambda_3}\omega_4^{i\lambda_4}
\mathcal{M}(2^-,3^+,4^+)_{\Phi*(Q)}\frac{\delta(Q^2)}{Q^2} \, , \label{eq:3ptbarMellin}
\end{align}
where
\be
\mathcal{M}(2^-,3^+,4^+)_{\Phi^*(Q)} =  \frac{ [34]^3}{[23]  [42] } = \frac{\omega_3\omega_4}{\omega_2} \frac{\bz_{34}^3}{\bz_{23}\bz_{42}}\,
\ee
is the
$\overline{\text{MHV}}$ amplitude evaluated in the antiholomorphic $\Phi^*$ dilaton background, see Eq.(\ref{mhvbd}).

After repeating the steps leading to Eq.(\ref{eq:3ptMellin}), we find that up to an overall normalization constant,
\begin{align}
\Big\langle\phi_{\D_2,-}^{a_2,-\epsilon}&(z_2,\bz_2)
\,\phi_{\D_3,+}^{a_3,+\epsilon}
(z_3,\bar z_3)
\,\phi_{\D_4,+}^{a_4,+\epsilon}(z_4,\bz_4)\Big\rangle_{\Phi^*}
 \nonumber\\
&=if^{a_2a_3a_4} \delta(\lambda_2+\lambda_3+\lambda_4) (1+i\lambda_3+i\lambda_4)B(-i\lambda_3,-i\lambda_4)\nonumber\\[1mm]
&~~~~~~\times z_{23}^{i\lambda_4}\bz_{23}^{i\lambda_4-1}z_{24}^{i\lambda_3} \bz_{24}^{i\lambda_3-1} z_{34}^{-i\lambda_3-i\lambda_4-2}  \bz_{34}^{-i\lambda_3-i\lambda_4+1}  \,.
\label{eq:3ptbarMellin}
\end{align}
Then the shadow transform (\ref{shad1}) can be performed by using the $n=3$ version of Osborn's conformal integral \cite{Osborn:2012vt}
\be
I_n={1\over \pi} \int\ d^2z\ f_n(z) \bar{f}_n(\bz), \quad f_n(z)= \prod_{i=1}^n {1\over (z-z_i)^{q_i}} \ , \quad \bar{f}_n(\bz)= \prod_{i=1}^n {1\over (\bz-\bz_i)^{\bar{q}_i}}
\ee
with
\be
\sum_{i=1}^n q_i=\sum_{i=1}^n \bar{q}_i=2 \ , \quad q_i-\bar{q}_i \in \mathbb{Z}\ .
\ee
By using Osborn's formulas, we find that indeed, up to an overall normalization constant,
\begin{align}\Big\langle\widetilde{\phi}_{\widetilde{\D}_2,+}^{a_2,-\epsilon}&(z_2,\bz_2)
\,\phi_{\D_3,+}^{a_3,+\epsilon}
(z_3,\bar z_3)
\,\phi_{\D_4,+}^{a_4,+\epsilon}(z_4,\bz_4)\Big\rangle_{\text{SV}} \nonumber\\ &=
\Big\langle\widetilde{\phi_{\D_2,-}^{a_2,-\epsilon}}(z_2,\bz_2)
\,\phi_{\D_3,+}^{a_3,+\epsilon}
(z_3,\bar z_3)
\,\phi_{\D_4,+}^{a_4,+\epsilon}(z_4,\bz_4)\Big\rangle_{\Phi^*}\ .
\end{align}
We conclude that the quasishadow correlator is identical to the shadow-transformed
$\overline{\text{MHV}}$ correlator evaluated in the antiholomorphic $\Phi^*$ dilaton background.
This indicates a deeper connection between shadows and quasishadows and deserves further study.
\subsection{MHV projection}
It is well known that the four-point correlators can be written as the sums of the so-called conformal blocks. Each block represents a ``conformal partial wave'' associated with a primary field propagating in a given two-dimensional channel \cite{Ferrara:1972kab,Ferrara:1972uq,Osborn:2012vt}. Here, we focus on the $s$-channel and on the primary fields with lowest Re$(\Delta)=1$, that is gluons and quasishadow fields appearing at the leading OPE order.
In order to construct the
respective partial wave, we start from
\begin{align}W_s[{\rm Re}(\D)=1]&=\sum_{a_p}
\int d\lambda_p\int d^2 z_p \Big\langle \phi_{\D_1,-}^{a_1,-\epsilon}(z_{1},\zbar_{1})\, \phi_{\D_2,-}^{a_2,-\epsilon}(z_2,\bz_2)\,\phi_{\D_p,+}^{a_p,+\epsilon}(z_p,\bz_p)\Big
\rangle_{\Phi} \nonumber\\ &~~~~~~~~~~~~~~~\times \Big\langle\phi_{2-\D_p,-}^{a_p,-\epsilon}(z_p,\bz_p)
\,\phi_{\D_3,+}^{a_3,+\epsilon}(z_3,\bar z_3)\,\phi_{\D_4,+}^{a_4,+\epsilon}(z_4,\bz_4)\Big\rangle_{\Phi^*}\ ,\label{pwave}
\end{align}
where $\D_p=1+i\lambda_p$. Note that one correlator originates from the MHV sector while the other one originates from $\overline{\text{MHV}}$;
therefore both sectors are treated symmetrically.  The three-point correlators are written in Eqs.(\ref{eq:3ptMHV}) and (\ref{eq:3ptbarMellin}):
\begin{align}
\Big\langle \phi_{\D_1,-}^{a_1,-\epsilon}&(z_{1},\zbar_{1})\, \phi_{\D_2,-}^{a_2,-\epsilon}(z_2,\bz_2)\,\phi_{\D_p,+}^{a_p,+\epsilon}(z_p,\bz_p)
\Big\rangle_{\Phi} \nonumber\\
&= \, i f^{a_1a_2 a_p}\, \delta(\lambda_1+\lambda_2+\lambda_p)\, (1+i\lambda_1+\lambda_2)B(-i\lambda_1,-i\lambda_2) \nonumber\\[1mm]
&~~~~~~~~~~\times z_{12}^{\, 1-i\lambda_1-i\lambda_2} \bz_{12}^{\, -i\lambda_1-i\lambda_2-2}\,z_{2p}^{\, i\lambda_1-1}\bz_{2p}^{\, i\lambda_1}\,z_{1p}^{\, i\lambda_2-1}  \bz_{1p}^{\, i\lambda_2} \, . \label{nonsh}
\end{align}
\begin{align}
\Big\langle\phi_{2-\D_p,-}^{a_p,-\epsilon}&(z_p,\bz_p)
\,\phi_{\D_3,+}^{a_3,+\epsilon}
(z_3,\bar z_3)
\,\phi_{\D_4,+}^{a_4,+\epsilon}(z_4,\bz_4)\Big\rangle_{\Phi^*}
 \nonumber\\
&=if^{a_pa_3a_4} \delta(\lambda_3+\lambda_4-\lambda_p) (1+i\lambda_3+i\lambda_4)B(-i\lambda_3,-i\lambda_4)\nonumber\\[1mm]
&~~~~~~\times z_{p3}^{i\lambda_4}\bz_{p3}^{i\lambda_4-1}z_{p4}^{i\lambda_3} \bz_{p4}^{i\lambda_3-1} z_{34}^{-i\lambda_3-i\lambda_4-2}  \bz_{34}^{-i\lambda_3-i\lambda_4+1}  \,.
\label{barsh}
\end{align}
In this way, we obtain
\begin{align}W_s[{\rm Re}(\D)=1]&=  f^{a_1a_2 b}  f^{a_3a_4b}
\delta(\lambda_1+\lambda_2+\lambda_3+\lambda_4)(1+i\lambda_1+\lambda_2)
(1+i\lambda_3+i\lambda_4)\nonumber\\ &~~~~~~~\times B(-i\lambda_1,-i\lambda_2) B(-i\lambda_3,-i\lambda_4) \, {\cal I}_p
\end{align}
where
\begin{align} {\cal I}_p&= \, z_{12}^{\, 1-i\lambda_1-i\lambda_2} \bz_{12}^{\, -i\lambda_1-i\lambda_2-2}
 z_{34}^{-i\lambda_3-i\lambda_4-2}\bz_{34}^{-i\lambda_3-i\lambda_4+1} \nonumber\\ &~~~~~~~~\times
\int d^2z_p\, z_{1p}^{\, i\lambda_2-1}\bz_{1p}^{\, i\lambda_2}  z_{2p}^{\, i\lambda_1-1} \bz_{2p}^{\, i\lambda_1} z_{p3}^{i\lambda_4} \bz_{p3}^{i\lambda_4-1}z_{p4}^{i\lambda_3} \bz_{p4}^{i\lambda_3-1} \end{align}
The above integral can be treated in the same way as in Ref.\cite{II} and reduced to
\begin{align}{\cal I}_p&=
 z_{12}^{~ -i \lambda_1- i\lambda_2+2}  \bar{z}_{12}^{~ -i \lambda_1- i\lambda_2-2}
  z_{13}^{~ -2-i\lambda_3}\bar{z}_{13}^{~-i\lambda_3}z_{14}^{~ -i\lambda_1-i\lambda_4}
 \bar{z}_{14}^{~ -i\lambda_1-i\lambda_4} z_{24}^{i\lambda_1-2} \bar{z}_{24}^{i\lambda_1}  \nonumber\\
&\times x^{-2-i\lambda_3-i\lambda_4} \bar{x}^{1-i\lambda_3-i\lambda_4} \int d^2 w \, w^{i\lambda_3} \bar{w}^{i\lambda_3-1} (w-1)^{i\lambda_1-1} (\bar{w}-1)^{i\lambda_1} (w-x)^{i\lambda_4}(\bar{w}-\bar{x})^{i\lambda_4-1} \ .
\end{align}
where, as usual, $x={z_{12}z_{34}}(z_{13}z_{24})^{-1}$ is the conformally invariant cross-ratio,
After using the formula written in Eq.(5.2) of Ref.\cite{II}, we obtain
\begin{align}W_s[{\rm Re}(\D)=1]&
= -\pi f^{a_1a_2 b}  f^{a_3a_4b}\, \delta(\lambda_1+\lambda_2+\lambda_3+\lambda_4)  \nonumber\\
&\times z_{12}^{~ -i \lambda_1- i\lambda_2+2}\bar{z}_{12}^{~ -i \lambda_1- i\lambda_2-2}
z_{13}^{~ -2-i\lambda_3}  \bar{z}_{13}^{~-i\lambda_3}z_{14}^{~ -i\lambda_1-i\lambda_4}\bar{z}_{14}^{~ -i\lambda_1-i\lambda_4}
z_{24}^{i\lambda_1-2}  \bar{z}_{24}^{i\lambda_1} \nonumber\\
&\times \Bigg[ (1+i\lambda_1+i\lambda_2)B(-i\lambda_1,-i\lambda_2)B(i\lambda_3,i\lambda_4)\nonumber\\[-2mm]
\label{eq:Duffintegral}
 &~~~~~~~~~\times \frac{1}{x} \, \, _2F_1\Big( {1-i\lambda_1,1+i\lambda_3\atop 2+i\lambda_3+i\lambda_4}; x\Big) \, _2F_1\Big( {-i\lambda_1,i\lambda_3\atop i\lambda_3+i\lambda_4}; \bar{x}\Big) \\[2mm]
&~~~~+ (1+i\lambda_3+i\lambda_4)B(i\lambda_1,i\lambda_2)B(-i\lambda_3,-i\lambda_4)\nonumber\\ &
~~~~~~~~~\times\frac{\bar{x}}{x^2}(x\bar{x})^{-i\lambda_3-i\lambda_4} \, _2F_1\Big({-i\lambda_4,i\lambda_2,\atop -i\lambda_3-i\lambda_4}; x\Big)\, _2F_1\Big({1-i\lambda_4,1+i\lambda_2,\atop 2-i\lambda_3-i\lambda_4}; \bar{x}\Big) \Bigg] \, . \nonumber
\end{align}

The conformal block decomposition of  four-point correlators will be discussed in full detail in the next section. Already at this point, however, we recognize the first term inside the square bracket in Eq.(\ref{eq:Duffintegral}) as the ${\rm Re}(\D)=1$, $J=1$ block in the decomposition of the MHV solution of BG equations. It agrees with the leading OPE term written in Eq.(\ref{eq:z34OPE}). The second term represents the ${\rm Re}(\D)=1$, $J=-1$ block which is absent in the $s$-channel decomposition of the MHV amplitude. This block appears though in the solution of
$\overline{\text{MHV}}$ BG equations, to be discussed below. It would imply an OPE different than in
Eq.(\ref{eq:z34OPE}): two positive helicity gluons would fuse into a negative helicity quasishadow field instead into a positive helicity gluon field.
Note that the difference between MHV and $\overline{\text{MHV}}$ blocks is in the spin of the block:
$W_s[{\rm Re}(\D)=1]|_{J=1}$ projects on the MHV block while $W_s[{\rm Re}(\D)=1]|_{J=-1}$ on
$\overline{\text{MHV}}$. In general, for blocks with higher ${\rm Re}(\D)$, $W_s|_{J>0}$ is the MHV projection while $W_s|_{J<0}$ is the $\overline{\text{MHV}}$ projection.

Four-gluon amplitudes are special because for a given MHV helicity configuration, they can be written in the momentum space either in the holomorphic MHV form or in the antiholomorphic $\overline{\text{MHV}}$ form. Before switching on the dilaton background, both forms are equivalent because of the momentum conservation law.\footnote{By using momentum conservation, one can also rewrite four-gluon amplitude in a ``hybrid'' form utilizing both holomorphic and antiholomorphic momentum spinors.}
When extrapolated to the amplitudes involving five or more gluons, the $\overline{\text{MHV}}$ formula
describes ``mostly minus'' helicity configurations. It is not difficult to derive  the $\overline{\text{MHV}}$ BG equations and apply them to exactly the same amplitude, as written in Eq.(\ref{partials}). One finds that the solutions of $\overline{\text{MHV}}$ equations are {\em different\/} from the solutions of MHV equations. In fact, $\overline{\text{MHV}}$ solutions can be obtained from MHV solutions (\ref{eq:sol_G1_sv},\ref{eq:sol_G2_sv}) by simply replacing $1\leftrightarrow 3, ~2\leftrightarrow 4$ and complex conjugating $z_i\leftrightarrow \bz_i$. The solutions of  $\overline{\text{MHV}}$  equations do {\em not\/} satisfy MHV equations though. They also yield different OPEs. At the leading order, the terms with antiholomorphic poles are exactly the same as in Refs.\cite{Fan1903,Strominger1910} but the terms with holomorphic poles are replaced by the quasishadow terms. The single-valued $\overline{\text{MHV}}$ solution can be constructed by summing over the partial waves, starting from Re$(\D)=1$ of Eq.(\ref{pwave}), and performing the $\overline{\text{MHV}}$ projection on the  primaries with $J<0$.

The origin of the difference between MHV and $\overline{\text{MHV}}$ solutions is clear. They correspond to different dilaton backgrounds. MHV corresponds to the holomorphic background of $\Phi$ while $\overline{\text{MHV}}$ corresponds to the antiholomorphic background of $\Phi^*$. One can ask if there exists a limit in which both solutions agree and eventually coincide with the distribution-valued celestial amplitude of Eqs.(\ref{pss1},\ref{pss2}). The answer is no. Even for $x=\bx$, we have two different backgrounds supplying momentum to the gluon system. The dilaton background cannot simply be switched off.

We end this section by adressing the important question of what is the complete set of CCFT operators.
It is well known that the partial waves can be constructed by using the shadow basis \cite{Osborn:2012vt}. In our case, we can write
\begin{align}W_s[{\rm Re}(\D)=1]&=\sum_{a_p}
\int d\lambda_p\int d^2 z_p \Big\langle \phi_{\D_1,-}^{a_1,-\epsilon}(z_{1},\zbar_{1})\, \phi_{\D_2,-}^{a_2,-\epsilon}(z_2,\bz_2)\,\widetilde\phi_{\widetilde\D_p,-}^{a_p,+\epsilon}(z_p,\bz_p)\Big
\rangle_{\Phi} \nonumber\\ &~~~~~~~~~~~~~~~\times \Big\langle
\widetilde{\phi}_{2-\widetilde{\D}_p,+}^{a_p,-\epsilon}(z_p,\bz_p)
\,\phi_{\D_3,+}^{a_3,+\epsilon}(z_3,\bar z_3)\,\phi_{\D_4,+}^{a_4,+\epsilon}(z_4,\bz_4)\Big\rangle_{\Phi^*}\ .\label{pwave1}
\end{align}
In standard CFT, this is just an alternative representation of the partial wave. In our case, however, it involves quasishadow fields that appear in the OPEs and should be included in the complete operator set. 

\section{Conformal block decomposition}
\subsection{$s$-channel}
In the $s$-channel [$x\approx 0, (12\rightleftharpoons 34)_{\bm{\mathfrak{2}}}$], the single-valued correlator (\ref{eq:full_MHV_SV}) can be decomposed into conformal blocks by applying the techniques developed in Refs.\cite{I,II}. They are based on Gau\ss\ recursion relations and on the channel decomposition of gauge group factors.
In the $s$-channel, the conformal block of a primary field with chiral weights $(h=\frac{\Delta+J}{2},\bar h=\frac{\Delta-J}{2})$ has the form \cite{Osborn:2012vt} :
\be K_{34}^{21}[h,\bh]=\bx^{\bh-\bh_3-\bh_4}\hy\left({\bh-\bh_{12},\bh+\bh_{34}\atop 2\bh};\bx\right)x^{h-h_3-h_4}\hy\left({h-h_{12},h+h_{34}\atop 2h};x\right)\ ,\label{blcs}\ee
where $h_{ij}=h_i-h_j$. In our case,
\begin{align}\nonumber&
h_{12}= \textstyle\frac{i\lambda_1}{2}-\textstyle \frac{i\lambda_2}{2}\ ,~  \bh_{12}=\frac{i\lambda_1}{2}-\textstyle\frac{i\lambda_2}{2},\qquad\qquad
h_{34}=\textstyle\frac{i\lambda_3}{2}-\textstyle\frac{i\lambda_4}{2}
\ ,~  \bh_{34}=\textstyle\frac{i\lambda_3}{2}-\textstyle\frac{i\lambda_4}{2}\ ,\\[1mm] &
h_3+h_4=2+\textstyle\frac{i\lambda_3}{2}+\textstyle\frac{i\lambda_4}{2}
\ ,~~~~~~~~~~~~~~~\bh_3+\bh_4=\textstyle\frac{i\lambda_3}{2}+\textstyle\frac{i\lambda_4}{2}\ .
\end{align}
After using Gau\ss\ recursion relations and basic properties of hypergeometric functions \cite{I,II}, we find that
\begin{align}
\mathcal{G}^{21}_{34}(x,\bar{x})&
\equiv f^{a_1a_2b}f^{a_3a_4b}G_{\text{SV}}(x,\bar{x})+f^{a_1a_3b}f^{a_2a_4b} \widetilde G_{\text{SV}}(x,\bar{x}) \nonumber\\[1mm]
&=\,\sum_{n=0}^\infty\sum_{J=1}^\infty (s_{n,J} f^{a_1a_2b}f^{a_3a_4b} +\tilde{s}_{n,J} f^{a_1a_3b}f^{a_2a_4b})\nonumber\\ &~~~~~~~~\times K^{21}_{34}\Big[ n+\frac{i\lambda_3}{2}+\frac{i\lambda_4}{2}+J, n+\frac{i\lambda_3}{2}+\frac{i\lambda_4}{2}\Big](x,\bar{x}) \, ,\nonumber\\[1mm]
&~~+ \sum_{n=0}^\infty\sum_{J=1}^\infty (t_{n,J} f^{a_1a_2b}f^{a_3a_4b} +\tilde{t}_{n,J} f^{a_1a_3b}f^{a_2a_4b})\nonumber\\&~~~~~~~\times K^{21}_{34}\Big[ 2+n-\frac{i\lambda_3}{2}-\frac{i\lambda_4}{2}+J, 2+n-\frac{i\lambda_3}{2}-\frac{i\lambda_4}{2}\Big](x,\bar{x}) \, ,\label{eq:s-chbl}
\end{align}
where
\begin{align}
s_{n,J} =& \textstyle (1+i\lambda_1+i\lambda_2-2n)B(n-i\lambda_1,n-i\lambda_2)B(n+i\lambda_3,n+i\lambda_4)\frac{\Gamma(i\lambda_3+i\lambda_4+2n)}{\Gamma(i\lambda_3+i\lambda_4+2n+2J-1)} \nonumber\\
&\times\textstyle \frac{1}{i\lambda_1+i\lambda_4}\Big( \frac{\Gamma(n+J-i\lambda_2)\Gamma(n+J+i\lambda_4)}{\Gamma(n-i\lambda_2)\Gamma(n+i\lambda_4)}- \frac{\Gamma(n+J-i\lambda_1)\Gamma(n+J+i\lambda_3)}{\Gamma(n-i\lambda_1)\Gamma(n+i\lambda_3)}\Big) \, , \\[3mm]
\tilde{s}_{n,J} =& -s_{n,J} -(-1)^J s_{n,J}(3\leftrightarrow4) \, , \\
&\,\nonumber \\
t_{n,J} =&\textstyle(i\lambda_3+i\lambda_4-3-2n)B(n+2+i\lambda_1,n+2+i\lambda_2)B(n+2-i\lambda_3,n+2-i\lambda_4) \nonumber\\
&\times \textstyle \frac{\Gamma(4-i\lambda_3-i\lambda_4+2n)}{\Gamma(3-i\lambda_3-i\lambda_4+2n+2J)} \frac{1}{i\lambda_2+i\lambda_3}\Big(\frac{\Gamma(n+J+2+i\lambda_1)\Gamma(n+J+2-i\lambda_3)}{\Gamma(n+2+i\lambda_1)\Gamma(n+2-i\lambda_3)}-\frac{\Gamma(n+J+2+i\lambda_2)\Gamma(n+J+2-i\lambda_4)}{\Gamma(n+2+i\lambda_2)\Gamma(n+2-i\lambda_4)} \Big) \, , \\
\tilde{t}_{n,J} = &-t_{n,J}-(-1)^J t_{n,J}(3\leftrightarrow4) \, .
\end{align}
The coefficients $s_{n,J}(3\leftrightarrow4)$ and $t_{n,J}(3\leftrightarrow4)$
are obtained from $s_{n,J}$ and $t_{n,J}$, respectively, by exchanging 3 and 4.

There are two sets of primaries  contributing in Eq.(\ref{eq:s-chbl}). The first set of primaries [from $a$ in Eq.(\ref{ssol})] has $\Delta_{a,n,J} = 2n+J+i\lambda_3+i\lambda_4$, with $n\geq 0$ and spin $J\geq 1$. The second set of primaries  [from $b$ in Eq.(\ref{ssol})] has $\Delta_{b,n,J} = 4+2n+J-i\lambda_3-i\lambda_4$, with $n\geq 0$ and spin $J\geq 1$.
They start at Re$(\Delta)=5$ and can be associated with the quasishadow fields discussed in the previous section.
Note that the limit of $\lambda_1=0$ is dominated by the $n=0$ block of the first set. The coefficients $s_{0,J}$ and $\tilde{s}_{0,J}$ match Eqs.(6.16) and (6.17) of Ref.\cite{II}.

In order to identify the group representations of conformal blocks, the color coefficients must be factorized into the  products of $s$-channel Clebsch-Gordan coefficients. In the simplest case of $SU(2)$, the biadjoint representation splits into three irreducible representations, with isospins 0, 1 and 2, and the group factors can be written as \cite{II}
\begin{align}
f^{a_1a_2b}f^{a_3a_4b} &= 2\sum_M C_{1M}^{a_1a_2}C_{1M}^{*\, a_3a_4}\, , \label{eq:f1s}\\
f^{a_1a_3b}f^{a_2a_4b} &= -\sum_M C_{2M}^{a_1a_2}C_{2M}^{*\, a_3a_4}+\sum_M C_{1M}^{a_1a_2}C_{1M}^{*\, a_3a_4}+2C_{00}^{a_1a_2}C_{00}^{* \, a_3a_4} \,  , \label{eq:f2s}\end{align}
where $C_{IM}^{ab}$ denote the Clebsh-Gordan coefficients for the fusion of two triplets into the states with isospin $I=0,1,2$.
Then
\begin{align}
s_{n,J} f^{a_1a_2b}&f^{a_3a_4b} +\tilde{s}_{n,J} f^{a_1a_3b}f^{a_2a_4b} \nonumber\\
=&\sum_M C_{1M}^{a_1a_2}C_{1M}^{*\, a_3a_4}\big[s_{n,J} -(-1)^J s_{n,J}(3\leftrightarrow4) \big] \nonumber\\
&+(2C_{00}^{a_1a_2}C_{00}^{* \, a_3a_4}-\sum_M C_{2M}^{a_1a_2}C_{2M}^{*\, a_3a_4})\big[ -s_{n,J} -(-1)^J s_{n,J}(3\leftrightarrow4)\big] \, , \label{eq:blcoeffs}
\end{align}
and a similar expression for other coefficients. For spin, $J=1$, only the $I=1$ triplet appears in the $s$-channel. For spin $J=2$ and higher, however, there are singlets, triplets, and quintuplets propagating in the $s$-channel.

It is interesting to check whether the block coefficients (\ref{eq:blcoeffs}) factorize (or not) into the (12) part multiplied by (34).
It is easy to check explicitly that the first few block coefficients do indeed factorize. For example, $n=0, J=1$ ; $n=0, J=2$ ; $n=1, J=1$; and $n=1, J=2$. For general $n$ and $J$, however, this property is not clear.
There are two cases, however, for general $n$, in which the coefficients factorize. When $J=1$, we obtain
\begin{align}
s_{n,1} f^{a_1a_2b}&f^{a_3a_4b} +\tilde{s}_{n,1} f^{a_1a_3b}f^{a_2a_4b} \nonumber\\
=& f^{a_1a_2b}f^{a_3a_4b}(1+i\lambda_1+i\lambda_2-2n)B(n-i\lambda_1,n-i\lambda_2)B(n+i\lambda_3,n+i\lambda_4) \, .
\end{align}
When $J=2$,
\begin{align}
&s_{n,2} f^{a_1a_2b}f^{a_3a_4b} +\tilde{s}_{n,2} f^{a_1a_3b}f^{a_2a_4b} \nonumber\\
&= C_{1M}^{a_1a_2}C_{1M}^{*\, a_3a_4}B(n-i\lambda_1,n-i\lambda_2)B(n+i\lambda_3,n+i\lambda_4)\frac{(i\lambda_1-i\lambda_2)(i\lambda_3-i\lambda_4)(2n-1+i\lambda_3+i\lambda_4)}{(2n+i\lambda_3+i\lambda_4)(2n+2+i\lambda_3+i\lambda_4)}  \nonumber\\
&+ (2C_{00}^{a_1a_2}C_{00}^{* \, a_3a_4}-\sum_M C_{2M}^{a_1a_2}C_{2M}^{*\, a_3a_4})(2n-1-i\lambda_1-i\lambda_2)B(n-i\lambda_1,n-i\lambda_2)B(n+i\lambda_3,n+i\lambda_4)\, .
\end{align}
{}For higher $J$, we found some cases when the coefficients do not factorize in a simple way. This may indicate certain degeneracy of the conformal block spectrum \cite{II}.
\subsection{$u$-  and $t$-channels}
In the $u$-channel $[x\approx 1, (14\rightleftharpoons 32)_{\bm{\mathfrak{2}}}]$, the conformal block of a primary field with chiral weights $(h,\bar h)$ has the form
\ba K_{32}^{41}&\!\![h,\bh](1-x,1-\bx)=(1-x)^{h-h_3-h_2}\hy\left(h-h_{14},h+h_{32}, \,2h; 1-x\right)\nonumber\\
&~~~~~~~~~~~~~~~\times(1-\bx)^{\bh-\bh_3-\bh_2}\hy\left(\bh-\bh_{14},\bh+\bh_{32},\, 2\bh;1-\bx\right)\ .\label{blsu}\ea
In our case
\begin{align}\nonumber&
h_{14}=h_{1}-h_4= \textstyle-1+\frac{i\lambda_1}{2}-\textstyle \frac{i\lambda_4}{2}\ , &\bh_{14}=\bh_{1}-\bar{h}_4=\textstyle1+\frac{i\lambda_1}{2}-\textstyle\frac{i\lambda_4}{2}\ ,\\[1mm] &
h_{32}=h_3-h_2=1+\textstyle\frac{i\lambda_3}{2}-\frac{i\lambda_2}{2}
\ , & \bh_{32}=\bh_3-\bh_2=-1+\textstyle\frac{i\lambda_3}{2}-\frac{i\lambda_2}{2}\ ,\\[1mm] &
h_3+h_2=1+\textstyle\frac{i\lambda_3}{2}+\textstyle\frac{i\lambda_2}{2}\ , & ~~~~~~\bh_3+\bh_2=1+\textstyle\frac{i\lambda_3}{2}+\textstyle\frac{i\lambda_2}{2}\ .\nonumber
\end{align}
By using Gau\ss\ recursion relations and the properties of hypergeometric functions, we can recast the $x\approx 1$ expansion of the single-valued correlator (\ref{eq:full_MHV_SV}) into the following sum of conformal blocks:
\begin{align}
\mathcal{G}_{34}^{21}(x,\bx) &=\mathcal{G}_{32}^{41}(1-x,1-\bx) \nonumber\\
=&\sum_{n=0}^{\infty}\sum_{J=-1}^{\infty}
(a_{n,J}\, f^{a_1a_2b}f^{a_3a_4b}+\tilde a_{n,J}\, f^{a_1a_3b}f^{a_2a_4b})\nonumber\\
&~~~\times
K_{32}^{41}{\Big[n+J+1+\frac{i\lambda_2}{2}+\frac{i\lambda_3}{2} ,\, n+1+\frac{i\lambda_2}{2}+\frac{i\lambda_3}{2}\Big](1-x,1-\bx)} \nonumber\\
~~+&\sum_{n=0}^{\infty}\sum_{J=-1}^{\infty}
(c_{n,J}\, f^{a_1a_2b}f^{a_3a_4b}+\tilde c_{n,J}\, f^{a_1a_3b}f^{a_2a_4b})\nonumber\\ &~~~\times
K_{32}^{41}{\Big[n+J+1-\frac{i\lambda_2}{2}-\frac{i\lambda_3}{2} ,\, n+1-\frac{i\lambda_2}{2}-\frac{i\lambda_3}{2}\Big](1-x,1-\bx)}  \, . \label{eq:u-chbl}
\end{align}

There are two sets of primaries contributing to Eq.(\ref{eq:u-chbl}). The first set of primaries has $\Delta_{ab,n,J} = 2n+J+2+i\lambda_2+i\lambda_3$, with $n\geq 0$ and spin $J\geq -1$. It contains, however, two subsets [from $a$ and $b$ in Eqs.(\ref{eq:solu_G1a}-\ref{eq:solu_G1d})], corresponding to incoming and outgoing fields, respectively,
with identical chiral weights.
This can be seen already at the leading OPE order, in the first two lines of Eq.(\ref{eq:z23OPE}).
 The second set has $\Delta_{cd,n,J} = 2n+J+2-i\lambda_2-i\lambda_3$  with $n\geq 0$ and spin $J\geq -1$ [from $c$ and $d$ in Eqs.(\ref{eq:solu_G1a}-\ref{eq:solu_G1d})]. It contains
the blocks associated with incoming and outgoing quasishadow fields, {\em c.f}.\ the last two lines of Eq.(\ref{eq:z23OPE}). Note that in the $u$-channel, the MHV projection selects the blocks with $J\ge -1$.

The computation of the conformal block coefficients in the $u$-channel is more difficult than in the $s$-channel because the relevant hypergeometric identities are more complicated. Nevertheless, they can be computed recursively. For $J=-1$ and general $n$, which include the blocks with lowest Re$(\Delta)=1 ~(n=0)$ and $J=-1$, we obtain
\begin{align}
&a_{n,-1}\, f^{a_1a_2b}f^{a_3a_4b}+\tilde a_{n,-1}\, f^{a_1a_3b}f^{a_2a_4b}=f^{a_1a_4b}f^{a_2a_3b} \nonumber\\
&\times\Big[ -B(i\lambda_3+n,-i\lambda_2-i\lambda_3-1-2n)B(n-i\lambda_1,i\lambda_4+i\lambda_1-1-2n)
(1-i\lambda_1-i\lambda_4+2n) \nonumber \\
&+B(n+2+i\lambda_2,-i\lambda_2-i\lambda_3-1-2n)
B(n+2-i\lambda_4,i\lambda_4+i\lambda_1-1-2n)(1+i\lambda_2+i\lambda_3+2n)\Big] , \\
&c_{n,-1}\, f^{a_1a_2b}f^{a_3a_4b}+\tilde c_{n,-1}\, f^{a_1a_3b}f^{a_2a_4b}=f^{a_1a_4b}f^{a_2a_3b} \nonumber\\
&\times\Big[ - B(i\lambda_4+n,-i\lambda_1-i\lambda_4-1-2n)B(n-i\lambda_2,i\lambda_3+i\lambda_2-1-2n)
(1-i\lambda_2-i\lambda_3+2n)\nonumber \\
&+ B(n+2+i\lambda_1,-i\lambda_1-i\lambda_4-1-2n)B(n+2-i\lambda_3,i\lambda_3+i\lambda_2-1-2n)
(1+i\lambda_1+i\lambda_4+2n)\Big] .
\end{align}
and similar expressions for higher $J$. All blocks with $J=-1$ are in the adjoint representation. Other representations, contained in the product of two adjoint representations, appear at $J>-1$.

Due to the symmetry of the single-valued correlator under $3\leftrightarrow 4$, which is equivalent to two-dimensional
$u\leftrightarrow t$ crossing symmetry,
the conformal block spectrum in the $t$-channel $[1/x\approx 0, (13\rightleftharpoons 42)_{\bm{\mathfrak{2}}}]$ is the same as in the $u$-channel, and the respective coefficients can be obtained from the $u$-channel by exchanging 3 and 4.
\section{Outlook}
Celestial conformal field theory is still at a very early stage of development. In this work, we discussed only some of its basic elements. We began by solving Banerjee-Ghosh equations for the four-point gluon correlators. By analyzing the solutions, we concluded that the action of two-dimensional CCFT must contain source terms similar to the background charge in Coulomb gas models and in Liouville theory. We identified these sources with the dilaton field on the celestial sphere. The dilaton background supplies momentum to the gluon system, hence allowing the construction of well-defined CFT correlators. We subjected the correlators to  standard scrutiny: extracted the OPEs, performed conformal block decomposition, determined the spectrum of primary fields and discussed crossing symmetry.

More detailed analysis of the correlation functions will certainly lead to a deeper understanding of CCFT. The goal is not only to construct a two-dimensional model of four-dimensional dynamics, but to use it in order to learn more about physics in four dimensions. This may happen in the following way. Most likely, the celestial theory extracted from tree-level Yang-Mills amplitudes will be a special limit of more general CFT. By relaxing such limits, we should be able to reach beyond the tree approximation, perhaps even beyond perturbation theory...

A very interesting problem to study is the backreaction of the dilaton background on four-dimensional asymptotically flat geometry. Can you see dilatons in the night sky? How do they affect gravity and cosmology?
\vskip 2mm
\noindent {\bf Acknowledgments}\\[2mm]
We are grateful to Sabrina Pasterski for organizing a wonderful workshop on ``Celestial Holography 2022'' at the Princeton Center for Theoretical Science, and for inviting us to present some preliminary results of this work. We thank Sabrina and the participants for asking many questions that helped us in preparing the present version. We are grateful to Shamik Banerjee and Sudip Ghosh for useful correspondence.
This material is based in part upon work supported by the National Science Foundation
under Grant Number PHY--1913328.
Any opinions, findings, and conclusions or recommendations
expressed in this material are those of the authors and do not necessarily
reflect the views of the National Science Foundation.
Wei Fan is supported in part by the National Natural Science Foundation of China under Grant No.\ 12105121.

\end{document}